\documentclass[12pt]{iopart}
\usepackage{bm}
\usepackage{graphics}
\usepackage{tikz}   

\newcommand{\be}{\begin{equation}}
\newcommand{\ee}{\end{equation}}
\newcommand{\ba}{\begin{eqnarray}}
\newcommand{\ea}{\end{eqnarray}}

\begin{document}

\title[Electromagnetic turbulence and global modes]{Gyrokinetic particle-in-cell simulations of electromagnetic turbulence in the presence of fast particles and global modes}

\author{A.~Mishchenko$^{(1)}$, A.~Bottino$^{(2)}$, T.~Hayward-Schneider$^{(2)}$, E.~Poli$^{(2)}$, X.~Wang$^{(2)}$, R.~Kleiber$^{(1)}$, M.~Borchardt$^{(1)}$, C.~{N\"uhrenberg}$^{(1)}$, A.~Biancalani$^{(3)}$, A.~{K\"onies}$^{(1)}$, E.~Lanti$^{(4)}$, Ph.~Lauber$^{(2)}$, R.~Hatzky$^{(2)}$, F.~Vannini$^{(2)}$, L.~Villard$^{(5)}$, and F.~Widmer$^{(6,2)}$}

\address{$^{(1)}$Max Planck Institute for Plasma Physics, D-17491 Greifswald, Germany\\
$^{(2)}$Max Planck Institute for Plasma Physics, D-85748 Garching, Germany \\
$^{(3)}$L\'eonard de Vinci P\^{o}le Universitaire, Research Center, 92 916 Paris La D\'efense, France\\
$^{(4)}$Ecole Polytechnique F\'ed\'erale de Lausanne, SCITAS, CH-1015 Lausanne, Switzerland\\
$^{(5)}$Ecole Polytechnique F\'ed\'erale de Lausanne, Swiss Plasma Center (SPC), CH-1015 Lausanne, Switzerland \\
$^{(6)}$International Research Collaboration Center, National Institutes of Natural Sciences, 105-0001 Tokyo, Japan}
\ead{alexey.mishchenko@ipp.mpg.de}
\vspace{10pt}
\begin{indented}
\item[]March 2022
\end{indented}

\begin{abstract}
Global simulations of electromagnetic turbulence, collisionless tearing modes, and Alfv\'en Eigenmodes in the presence of fast particles are carried out using the gyrokinetic particle-in-cell codes ORB5 (E.~Lanti~et~al, Comp.~Phys.~Comm, {\bf 251}, 107072 (2020)) and EUTERPE (V.~Kornilov~et~al, Phys. Plasmas, 11, 3196 (2004)) in tokamak and stellarator geometries. Computational feasibility of simulating such complex coupled systems is demonstrated. 
\end{abstract}

%
\section{Introduction}

Creation and control of burning plasmas is an ultimate goal of the magnetic fusion world-wide effort. Such plasmas will become experimentally accessible in the foreseeable future when the ITER and SPARC facilities start their operation. One of the characteristic features of the burning plasmas is the intrinsic richness of their physics featuring complex couplings and interactions of the microscopic processes (turbulence) with the macroscopic ones (MHD and Alfv\'enic modes). These interactions can affect properties of the turbulence as well as the evolution of macroscopic instabilities. Such couplings may become especially strong in burning plasmas where fast particles are abundant since they can drive the macroscopic modes unstable. A global approach is needed to assess the physics combining macroscopic modes and turbulence.

Global simulations of the electromagnetic turbulence can be particularly difficult due to the variety of the physics involved, but also as a consequence of the numerical problems, such as the cancellation problem \cite{YangChen,Mishchenko_mitigation}. Involving all the relevant time scales ranging from the electron motion to slow MHD dynamics (e.g. growth and nonlinear evolution of the tearing instability) and electromagnetic turbulence adds to the computational difficulty of the global gyrokinetic simulations in the electromagnetic regime. Addressing this problem in its full complexity for the reactor-scale plasma will require the emerging exascale computing. However, first steps in this direction can already be made now using the existing high-performing computing systems and employing the available codes. This has been the main goal of the PRACE computing project EMGKPIC. In this paper we report on the basic approach and the main results of the project. Because of the limited space, the results will be described in a brief manner leaving details to follow-up publications.

In the project, the global gyrokinetic particle-in-cell codes ORB5 \cite{Lanti} and EUTERPE \cite{Kornilov04} have been used to simulate electromagnetic turbulence in realistic tokamak and stellarator geometries. Demonstrating feasibility of global turbulence simulations using the electromagnetic gyrokinetic PIC codes was the first goal of the project. The global setup permits combining electromagnetic turbulence with global modes, such as tearing instabilities and Alfv\'enic Eigenmodes in the presence of the energetic particles. Assessing such combinations and affordability of their simulations was another goal of the project. A particularly strong emphasis has been put on the simulations of electromagnetic turbulence which is known to be notoriously challenging in a global setup. Combining it with the macroscopic physics (Alfv\'enic and MHD) is less difficult if the turbulence problem is solved. The following ``case studies'' have been identified:
\begin{enumerate}
	\item global electromagnetic turbulence in tokamak geometry (e.~g.~ASDEX-Upgrade)
	\item tearing instabilities and their combination with the electromagnetic turbulence
	\item nonlinear Alfv\'enic modes in the presence of fast particles and their combination with the electromagnetic turbulence
	\item global electromagnetic turbulence in stellarator geometry (Wendelstein 7-X)
\end{enumerate}
The ORB5 and EUTERPE codes share the equations solved, the basic discretization principles, and many aspects of technical implementation (see Ref.~\cite{Mishchenko_2021} for details). Despite these similarities, they remain separate projects with a different set of capabilities. Thus, it is only EUTERPE which can address stellarator geometries. On the other hand, tokamak simulations using ORB5 are normally more efficient due to the tokamak axisymmetry explicitly employed in the code. Therefore in the project, all tokamak simulations have been performed using ORB5 and all stellarator simulations using EUTERPE.

In this paper, we report global electromagnetic simulations solving gyrokinetic equations and employing the numerical schemes described in detail in Ref.~\cite{Mishchenko_2021}. According to the "case studies" selected and described above, the first basic component considered in these simulations is electromagnetic turbulence nonlinearly evolving and saturating through zonal flow excitation or relaxation of the plasma profiles. In the finite-beta regime, the Kinetic Ballooning Mode (KBM) turbulence is believed to play an important role \cite{pueschel_kbm,Ishizawa2015,Ishizawa2019,Maeyama2014,Kumar_2021}. We identify the KBM regime performing parameter scans with respect to the plasma beta. We compare the low-beta Ion-Temperature-Gradient driven (ITG) turbulence with the KBM turbulence at a higher beta in the same magnetic configuration and for the same plasma profile shape.
The second basic component of the global physics considered here is an MHD perturbation, namely the collisionless tearing mode, coupled to electromagnetic turbulence. We provide an example of this instability evolving separately and in the presence of the background turbulence. A basic feasibility of such simulations is demonstrated. 
The third component of the global physics is associated with macroscopic nonlinear Alfv\'enic Eigenmodes destabilized by the fast particles. Frequency evolution of a Toroidal Alfv\'en Eigenmode in the presence of electromagnetic turbulence is considered. 
Finally, the electromagnetic turbulence in non-axisymmetric stellarator geometry is addressed and feasibility of stellarator simulations is demonstrated. 
These examples encompass the physics content computationally available for future in-depth studies using ORB5 and EUTERPE.

The structure of this paper is as follows. In Sec.~\ref{Turbulence}, we consider electromagnetic turbulence in tokamak plasmas. In Sec.~\ref{Tearing}, the evolution of the tearing mode in the presence of electromagnetic turbulence is addressed. In Sec.~\ref{Fast_ions}, the coupled nonlinear system including Alfv\'en Eigenmodes, fast particles, and turbulence is considered. In Sec.~\ref{Stellarators}, the stellarator simulations are described. Finally, conclusions are made in Sec.~\ref{Conclusions}.
%
%
\section{Electromagnetic turbulence in tokamak plasmas}  \label{Turbulence}
%

\begin{figure}
\includegraphics[width=0.47\textwidth]{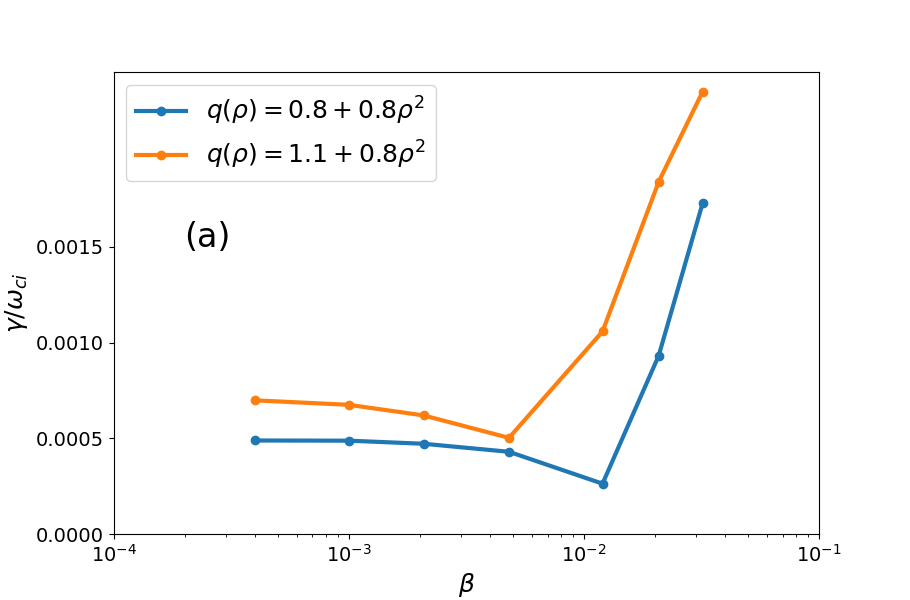}
\includegraphics[width=0.47\textwidth]{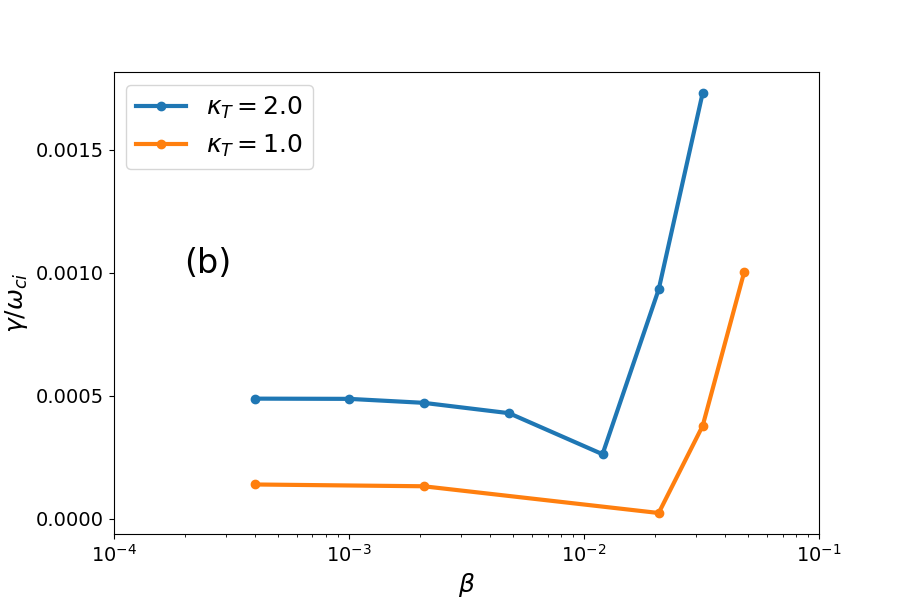}
\caption{Growth rate computed from the linear phase of the heat flux evolution plotted as a function of $\beta$. One clearly sees the characteristic ITG-KBM transition. (a) Temperature and density gradients $\kappa_{{\rm T}} = 2.0$, $\kappa_{{\rm n}} = 0.3$, safety factor profile $q(\rho)= 0.8 + 0.8 \rho^2$ compared to $q(\rho) = 1.1 + 0.8 \rho^2$. (b) Density gradient $\kappa_{{\rm n}} = 0.3$, safety factor $q(\rho)= 0.8 + 0.8 \rho^2$, temperature gradient $\kappa_{{\rm T}} = 2.0$ compared to $\kappa_{{\rm T}} = 1.0$.}
\label{gamma-beta}
\end{figure}

\begin{figure}
\includegraphics[width=0.47\textwidth]{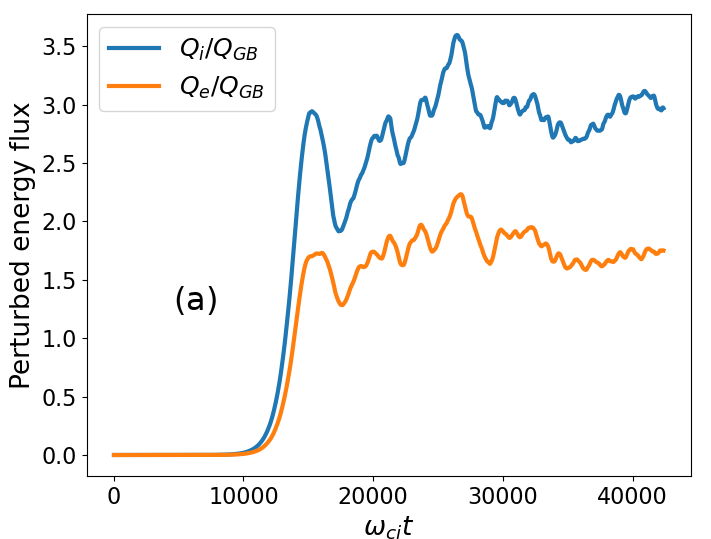}
\includegraphics[width=0.47\textwidth]{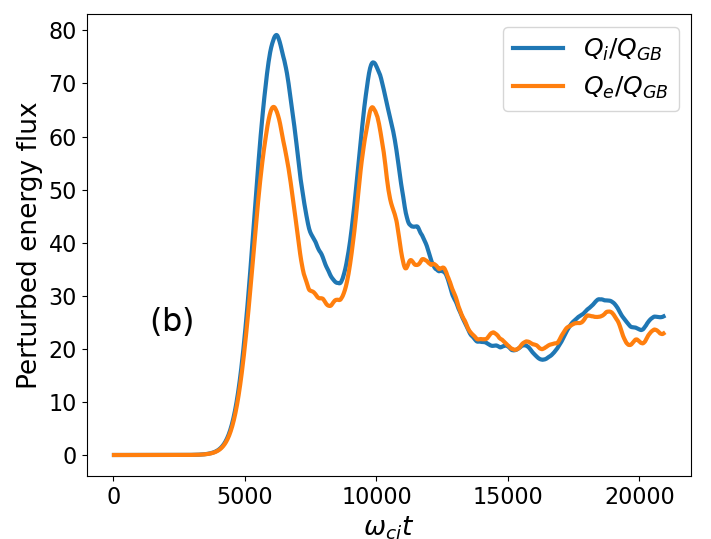}
\caption{Perturbed energy flux evolution for the safety factor $q(\rho) = 1.1 + 0.8 \rho^2$, the temperature gradient $\kappa_{{\rm T}}=2.0$, (a) $\beta = 0.1\%$ (electromagnetic ITG regime), and (b) $\beta = 2.08\%$ (KBM regime). The energy flux is considerably larger in the KBM regime.}
\label{hflux_t_ITG-KBM}
\end{figure}

First, we consider tokamak geometry with the aspect ratio $A = 10$, cirular cross-sections, the safety factors $q(\rho) = 0.8 + 0.8 \rho^2$ and $q(\rho) = 1.1 + 0.8 \rho^2$ (two cases compared), where $\rho$ is the radius of the circular flux surface, and the temperature and density profiles:
\ba
\label{dens_prof}
&&{} n_{0s}(s)/n_{0s}(s_0) = \exp\left[-\kappa_{{\rm n}} \Delta_{{\rm n}} {\rm tanh}\left(\frac{s - s_0}{\Delta_{{\rm n}}}\right)\right] \\
\label{temp_prof}
&&{} T_{0s}(s)/T_{0s}(s_0) = \exp\left[-\kappa_{{\rm T}} \Delta_{{\rm T}} {\rm tanh}\left(\frac{s - s_0}{\Delta_{{\rm T}}}\right)\right]
\ea
Here, $s = \sqrt{\psi/\psi_a}$, $\psi$ is the poloidal magnetic flux, $\psi_a$ is the poloidal magnetic flux at the plasma edge, $s_0 = 0.5$, $\kappa_{{\rm n}} = 0.3$, and $\Delta_{{\rm T}} = \Delta_{{\rm n}} = 0.208$. Two different temperature gradients will be considered corresponding to $\kappa_{{\rm T}} = 1.0$ and $\kappa_{{\rm T}} = 2.0$. The ubiquitous nonlinear generation of the small scales in the phase space (filamentation) is controlled with the Krook operator, see Ref.~\cite{Biancalani} for further details. The machine size is determined by $L_x = 2 r_{{\rm a}} / \rho_{{\rm s}} = 360$ with $r_{{\rm a}}$ the minor radius and $\rho_{{\rm s}}$ the characteristic bulk-ion gyroradius, the ion-to-electron mass ratio is $m_{{\rm i}}/m_{{\rm e}} = 200$. The character of the turbulence is defined by the plasma $\beta$: the electromagnetic-ITG regime for small $\beta$ changing into the KBM regime when $\beta$ increases. This dependence is shown in Fig.~\ref{gamma-beta} where the linear growth rate is plotted as a function of $\beta$. The growth rate is computed from the linear evolution of the perturbed heat flux. One sees a characteristic ITG-KBM transition for different plasma temperature and safety factor profiles. The nonlinear evolution of the radial heat flux is shown in Fig.~\ref{hflux_t_ITG-KBM} for the both regimes.

\begin{figure}
\includegraphics[width=0.32\textwidth]{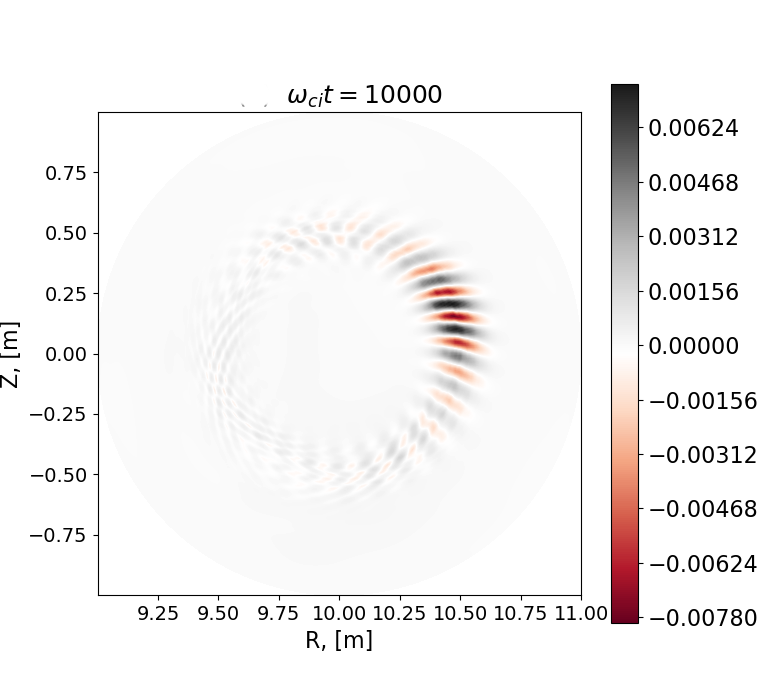}
\includegraphics[width=0.32\textwidth]{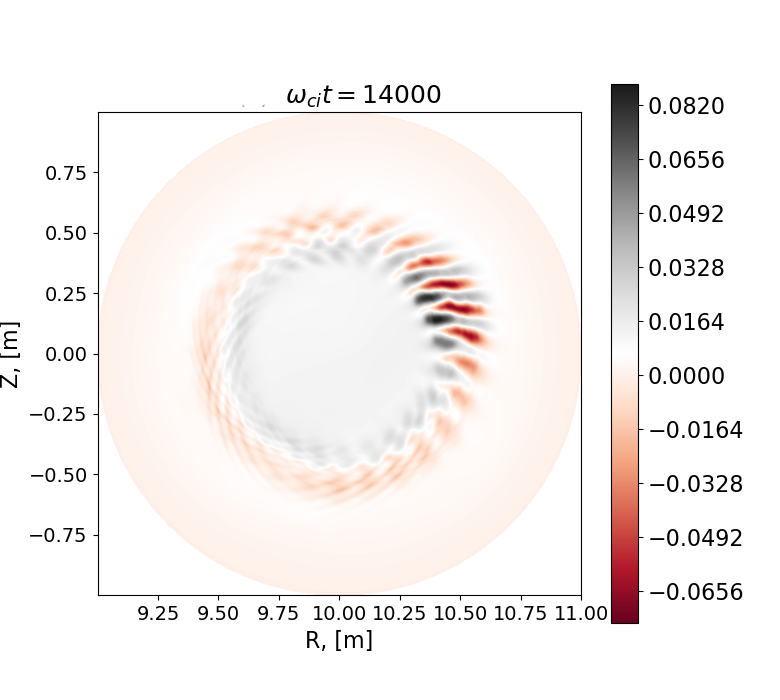}
\includegraphics[width=0.32\textwidth]{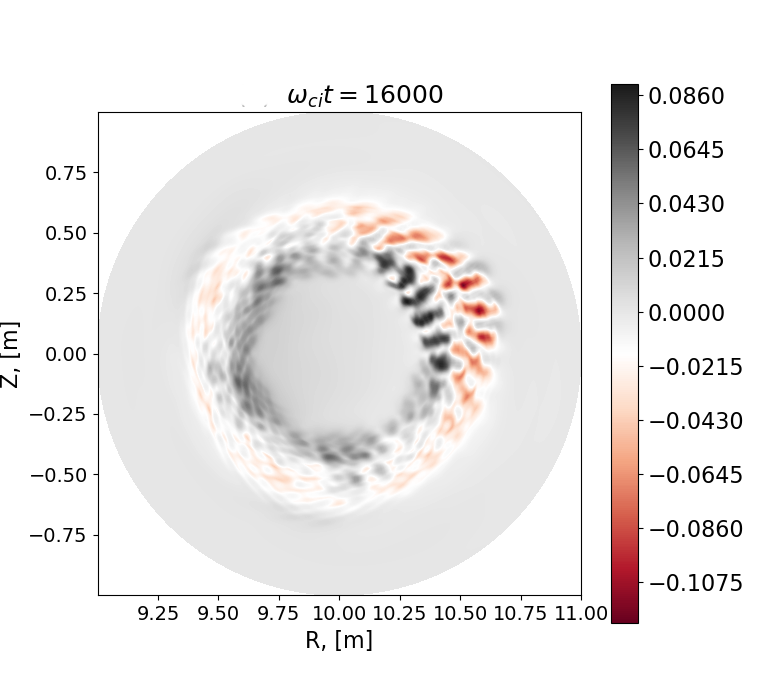}
\caption{Evolution of the electrostatic potential in the ITG turbulence shown in Fig.~\ref{hflux_t_ITG-KBM}(a). One sees how the zonal flow develops and decorrelates the turbulent eddies. Zonal flow dynamics determines the saturation of the electromagnetic ITG turbulence.}
\label{potsc_qa1.9-ITG}
\end{figure}

\begin{figure}
\includegraphics[width=0.32\textwidth]{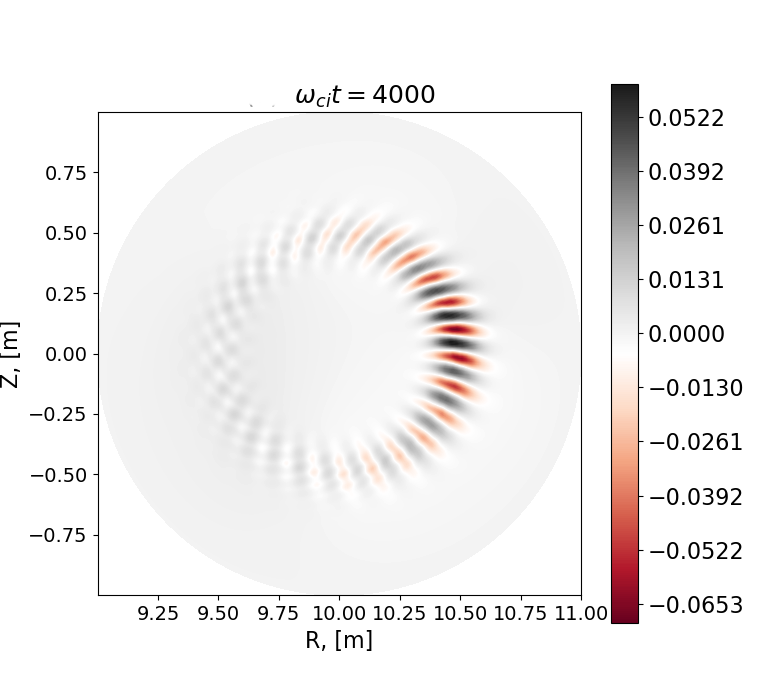}
\includegraphics[width=0.32\textwidth]{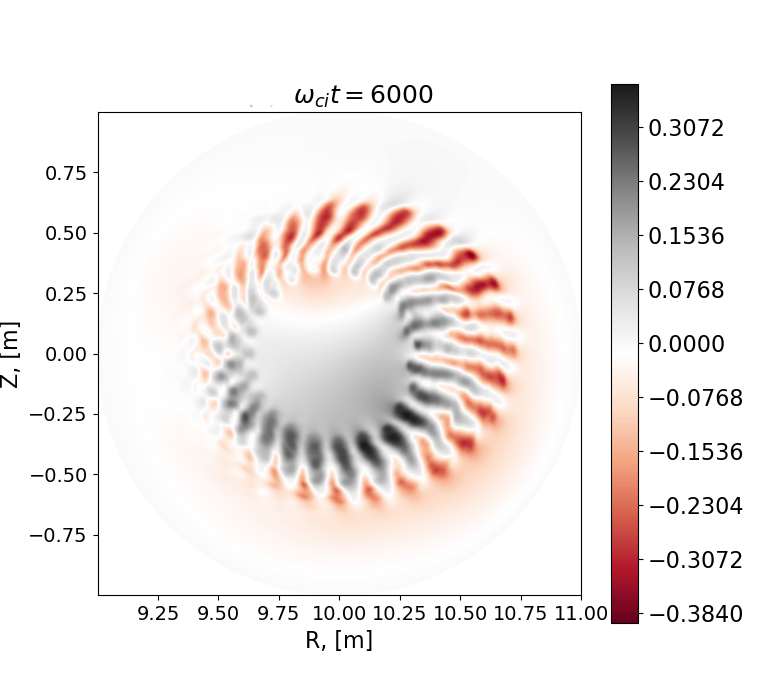}
\includegraphics[width=0.32\textwidth]{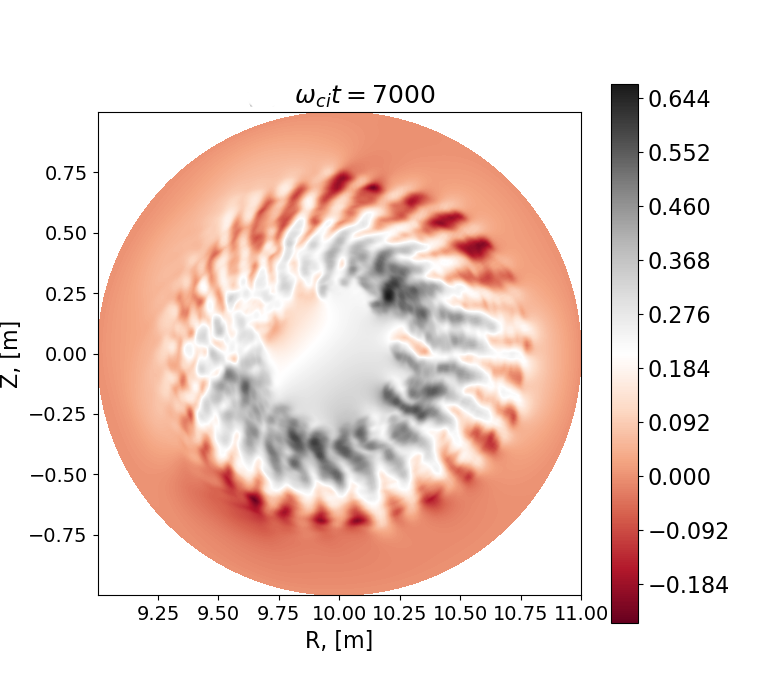}
\caption{Evolution of the electrostatic potential in the KBM turbulence shown in Fig.~\ref{hflux_t_ITG-KBM}(b). One sees the finger-like structures developing and propagating outwards. Plasma profile relaxation, Fig.~\ref{temp_relaxed_qa1.9}, determines the saturation of the KBM turbulence.}
\label{potsc_qa1.9}
\end{figure}

\begin{figure}
\includegraphics[width=0.47\textwidth]{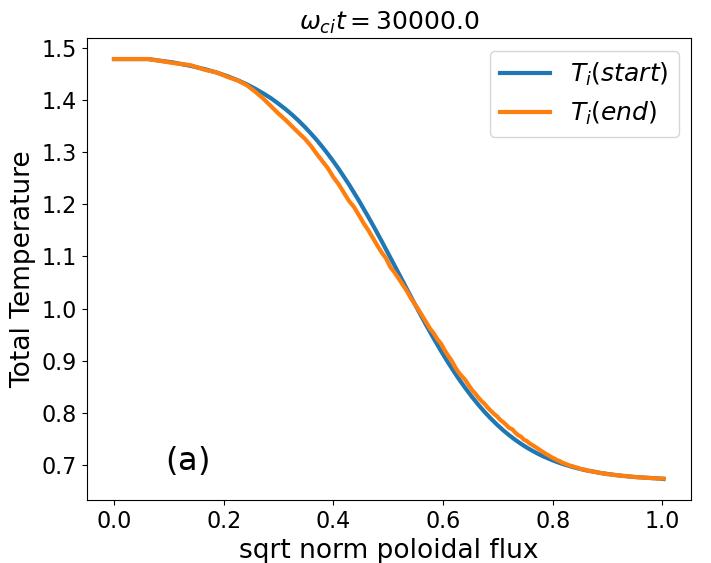}
\includegraphics[width=0.47\textwidth]{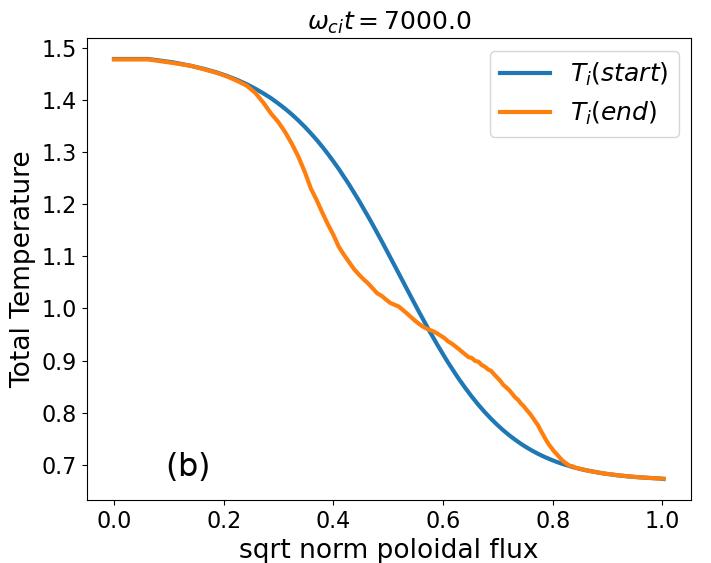}
\caption{Temperature relaxation in (a) the ITG regime ($\beta = 0.1\%$, $k_T = 2.0$, $q(\rho) = 1.1 + 0.8 \rho^2$), and (b) the KBM regime ($\beta = 2.08\%$, $k_T = 2.0$, $q(\rho) = 1.1 + 0.8 \rho^2$). The profile relaxation is much weaker for the ITG turbulence than in the KBM case.}
\label{temp_relaxed_qa1.9}
\end{figure}

\begin{figure}
	\centering
	\includegraphics[width=0.47\textwidth]{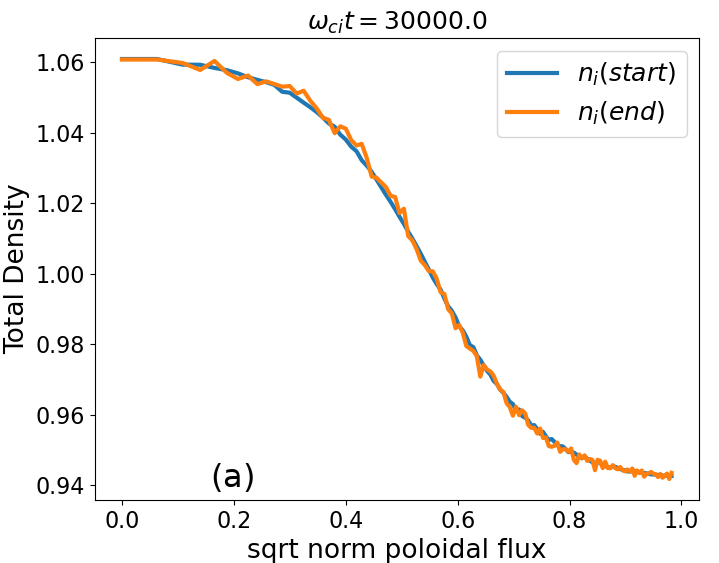}
	\includegraphics[width=0.47\textwidth]{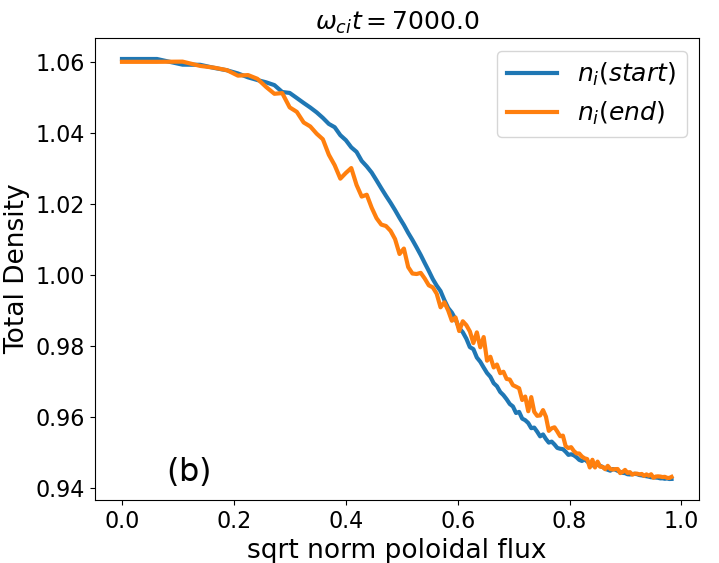}
	\caption{The density profile relaxation for (a) the ITG turbulence and (b) the KBM turbulence. The simulation parameters are as in Fig.~\ref{temp_relaxed_qa1.9}.}
	\label{dens_relaxed_qa1.9}
\end{figure}

In Fig.~\ref{hflux_t_ITG-KBM}, one sees that the simulation enters the nonlinear phase after the linear growth in both (a) the ITG regime and (b) the KBM regime. The latter case shows an oscillatory relaxation dynamics resulting from the combination of the profile flattening by the turbulence and an energy source, provided here by the Krook operator. This profile relaxation process is also indicated by the evolution of the electrostatic potential shown in Fig.~\ref{potsc_qa1.9}. One sees here how the finger-like structures develop and propagate outwards resulting in a flattening of the temperature profile shown in Fig.~\ref{temp_relaxed_qa1.9}(b). At later times, the temperature gradient restores to some extent leading to the next relaxation cycle. In contrast, the ITG regime shows a much weaker relaxation of the temperature profile, see Fig.~\ref{temp_relaxed_qa1.9}(a). The electrostatic potential, shown in Fig.~\ref{potsc_qa1.9-ITG}, develops the characteristic pattern of the turbulent eddies decorrelated by the zonal flow. The evolution of the density profile is shown in Fig.~\ref{dens_relaxed_qa1.9}.


\begin{figure}
	\centering
    \includegraphics[width=0.47\textwidth]{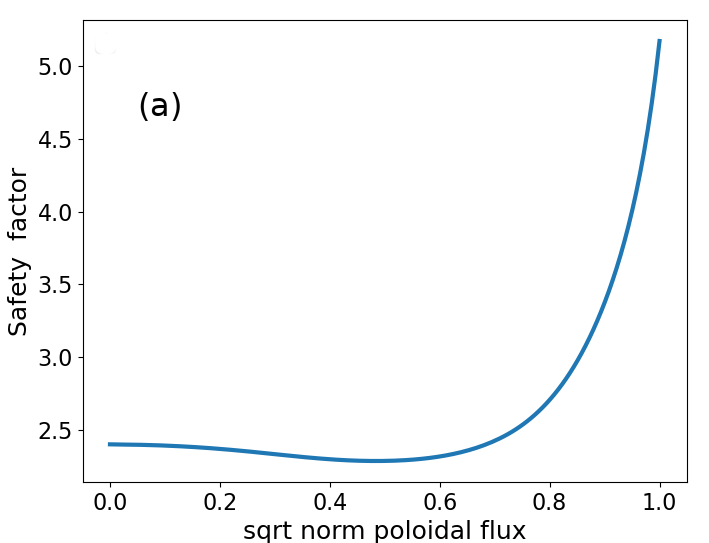}
    \includegraphics[width=0.47\textwidth]{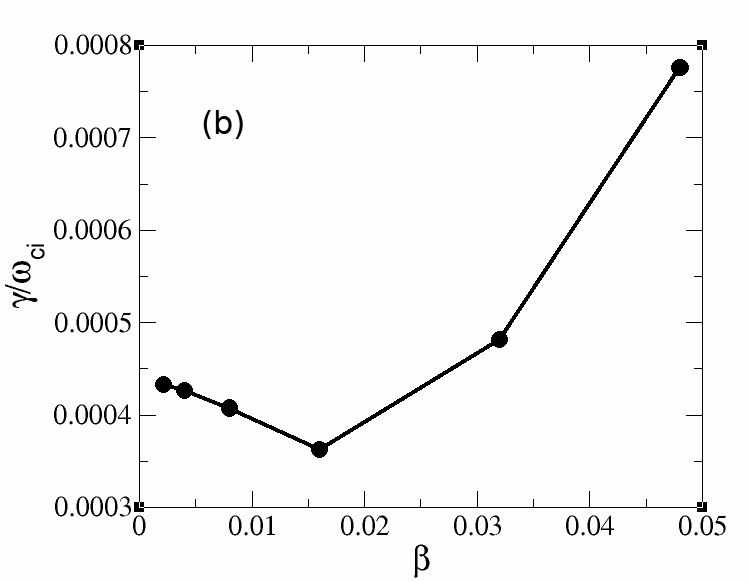}	
\caption{(a) The safety factor of the ``NLED-AUG'' case \cite{Vannini_2021}. (b) The growth rate in the ASDEX-Upgrade plotted as a function of plasma $\beta$. One can see the ITG-KBM transition similar to the circular case.}
	\label{asdex:gammabeta}
\end{figure}
\begin{figure}
	\centering
	\includegraphics[width=0.47\textwidth]{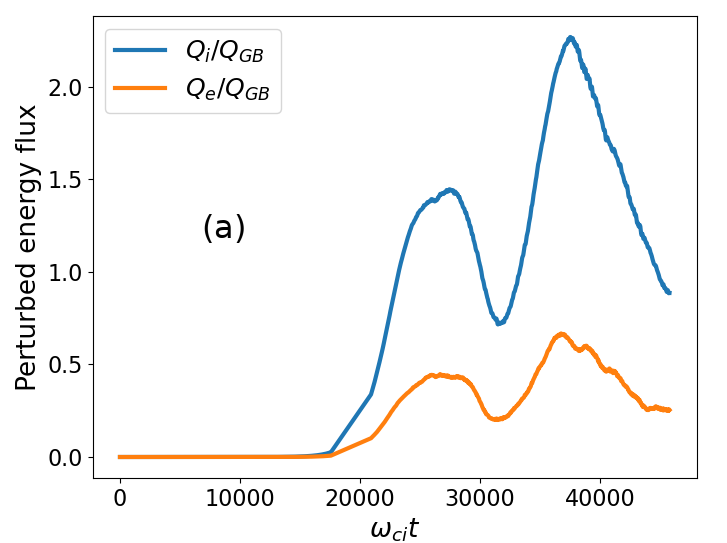}
	\includegraphics[width=0.47\textwidth]{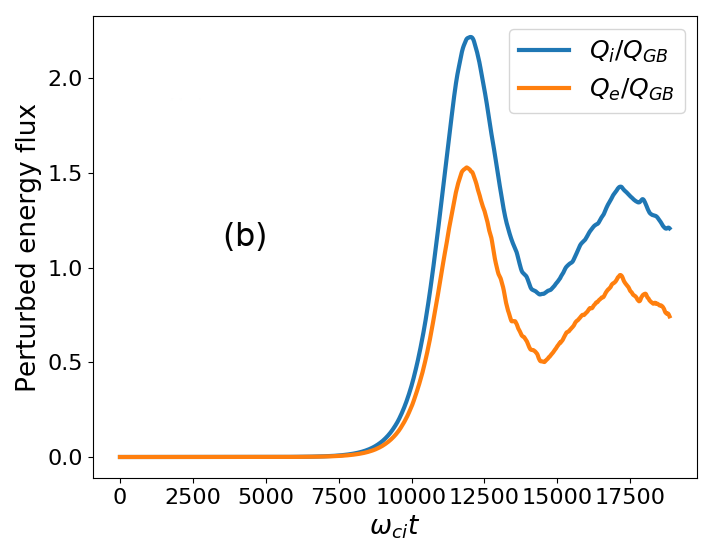}
	\caption{The perturbed energy flux in the ASDEX-Upgrade for (a) the ITG regime, $\beta = 0.4\%$, and (b) the KBM regime, $\beta = 4.8\%$.}
	\label{asdex:eflux}
\end{figure}
\begin{figure}
	\centering
	\includegraphics[width=0.32\textwidth]{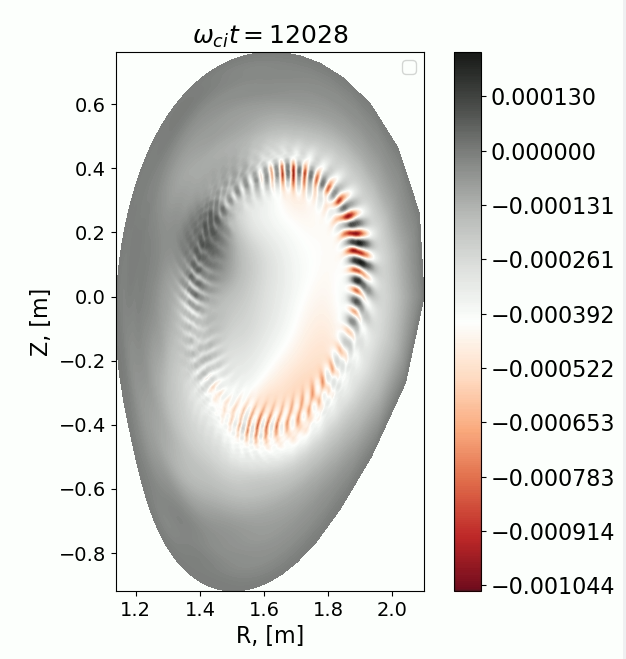}
	\includegraphics[width=0.32\textwidth]{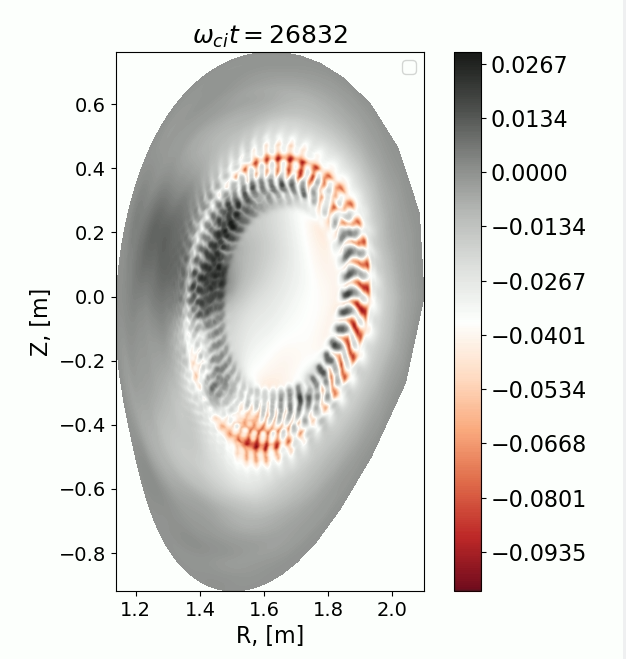}
	\includegraphics[width=0.32\textwidth]{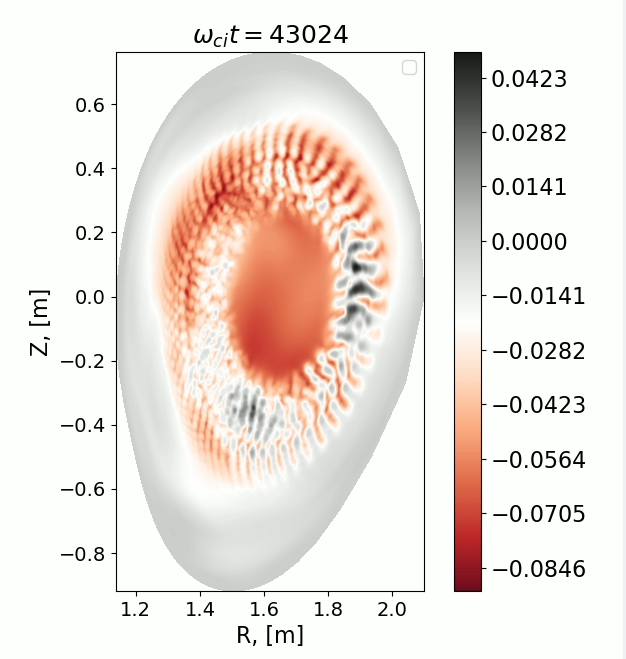}
	\caption{Evolution of the electrostatic potential in ASDEX-Upgrade for the ITG regime, $\beta = 0.4\%$.}
	\label{asdex:potsc_b0.001}
\end{figure}
\begin{figure}
	\centering
	\includegraphics[width=0.32\textwidth]{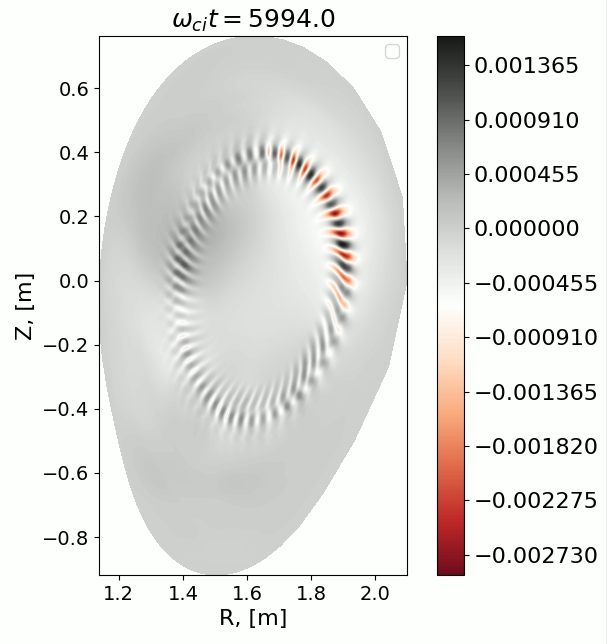}
	\includegraphics[width=0.32\textwidth]{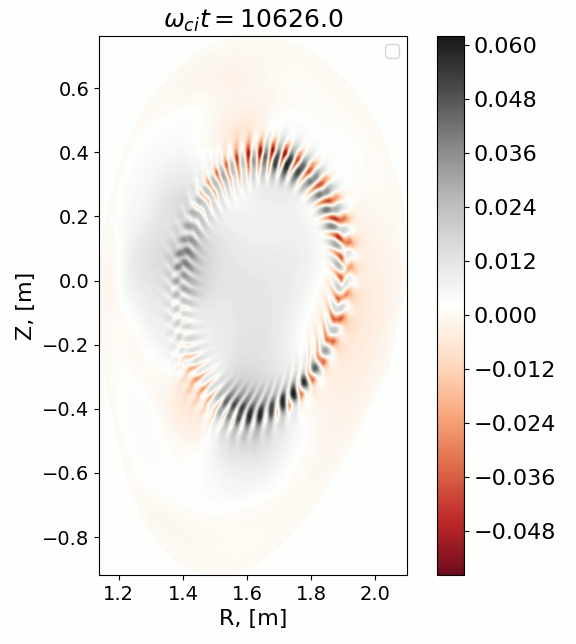}
	\includegraphics[width=0.32\textwidth]{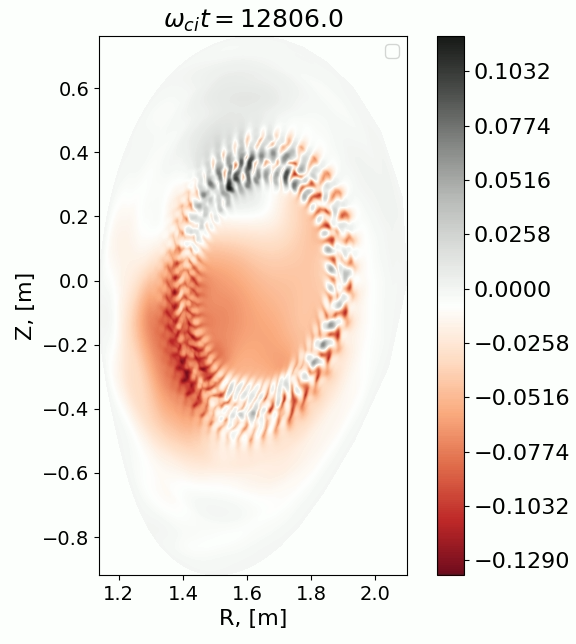}
	\caption{Evolution of the electrostatic potential in ASDEX-Upgrade for the KBM regime, $\beta = 4.8\%$.}
	\label{asdex:potsc_b0.012}
\end{figure}
\begin{figure}
	\centering
	\includegraphics[width=0.47\textwidth]{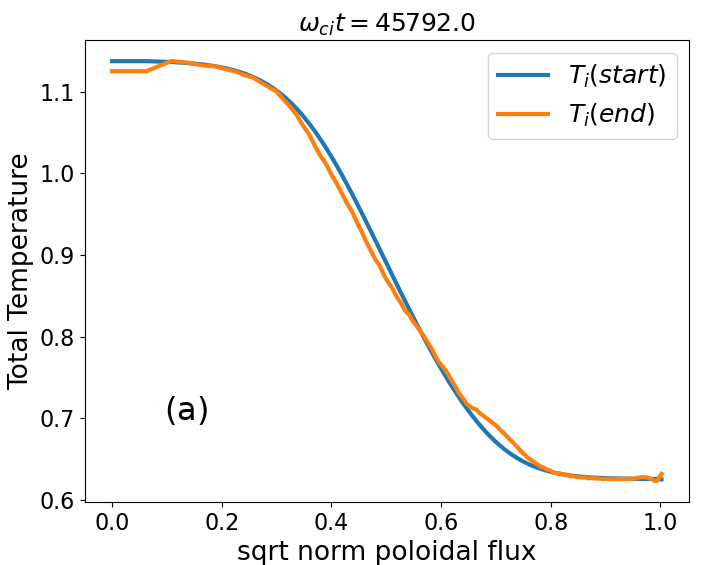}
	\includegraphics[width=0.47\textwidth]{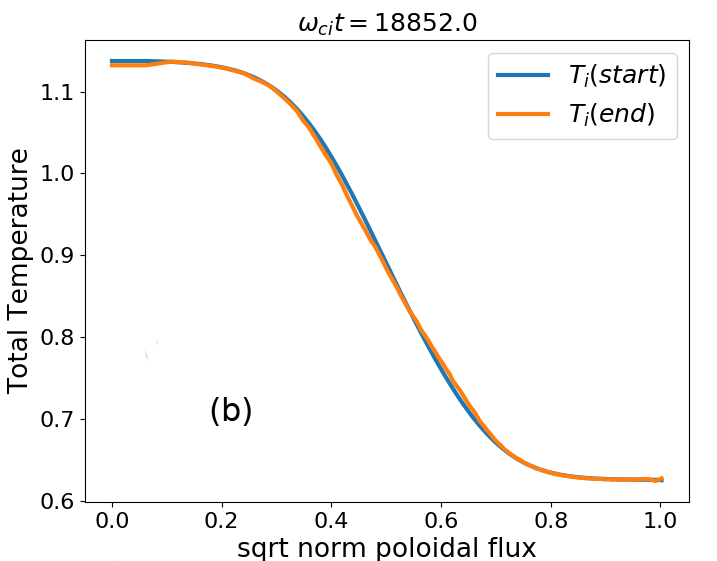}
	\caption{Temperature profile evolution in the ASDEX-Upgrade geometry in (a) the ITG regime, $\beta = 0.4\%$, and (b) the KBM regime, $\beta = 4.8\%$. One sees that the profiles do not change much in contrast to the more unstable circular-geometry case.}
	\label{asdex:temprelaxed}
\end{figure}
\begin{figure}
	\centering
	\includegraphics[width=0.47\textwidth]{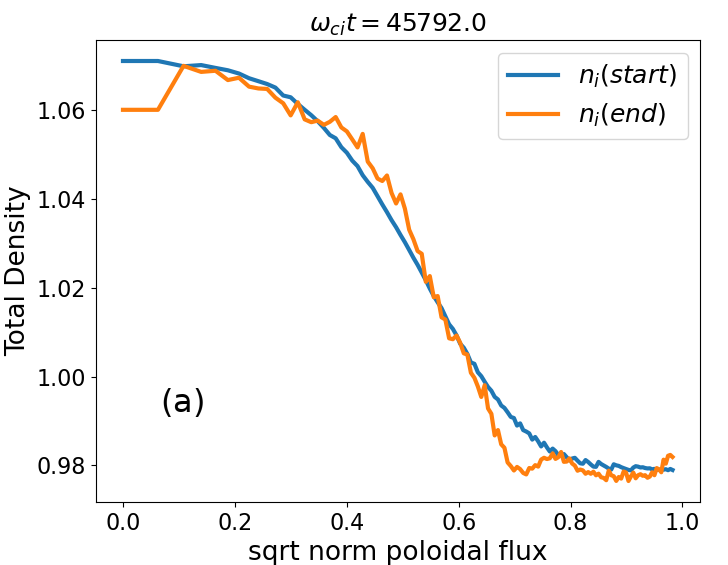}
	\includegraphics[width=0.47\textwidth]{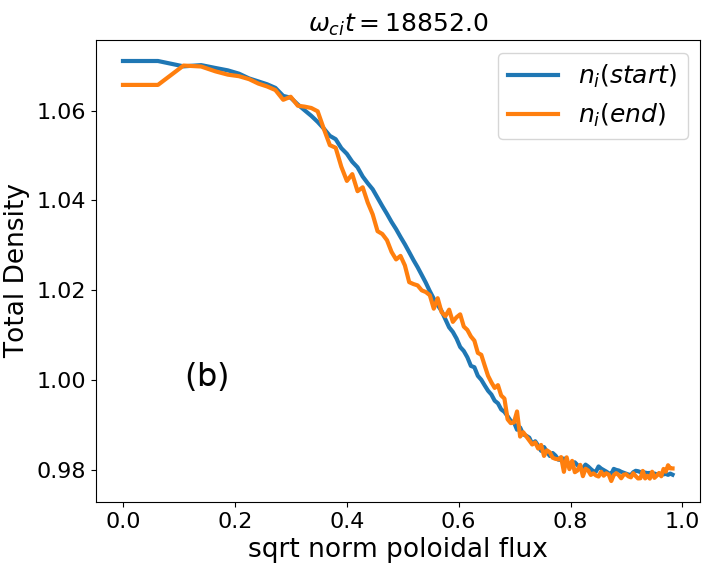}
	\caption{Density profile evolution for (a) the ITG regime, $\beta = 0.4\%$, and (b) the KBM regime, $\beta = 4.8\%$. One can see a density peaking (corresponding to an inward turbulent pinch) in the ITG regime. The particle flux in the KBM regime is outward.}
	\label{asdex:densrelaxed}
\end{figure}

As a next step, consider a realistic ASDEX-Upgrade equilibrium, similar to the ``NLED-AUG'' case \cite{Vannini_2021}, down-scaled to $L_x = 300$, with the safety factor profile shown in Fig.~\ref{asdex:gammabeta}(a); temperature and density profiles shown in Figs.~\ref{asdex:temprelaxed} and \ref{asdex:densrelaxed}. Similar to the previous case, the ITG-KBM transition can be seen in Fig.~\ref{asdex:gammabeta}(b) where the growth rate is shown as a function of plasma $\beta$ computed using the linear phase of the perturbed energy flux evolution. The perturbed energy flux is shown in Fig.~\ref{asdex:eflux}(a) for the ITG regime and in Fig.~\ref{asdex:eflux}(b) for the KBM regime. One can see that the energy flux has entered the nonlinear phase of its evolution. The values of the flux measured in the gyro-Bohm units are comparable in both regimes. The evolution of the electrostatic potential is shown in Fig.~\ref{asdex:potsc_b0.001} for the ITG regime, and in Fig.~\ref{asdex:potsc_b0.012} for the KBM regime. The temperature and density nonlinear profile evolution is plotted in Figs.~\ref{asdex:temprelaxed} and \ref{asdex:densrelaxed}. One can see that the temperature profile relaxation caused by the KBM turbulence is not very strong for the shaped ASDEX-Upgrade geometry, in contrast to the circular cross-section tokamak case shown in Fig.~\ref{temp_relaxed_qa1.9}(b). In future, the effect of the plasma shaping on the electromagnetic turbulence in tokamaks will be addressed in more detail. In Fig.~\ref{asdex:densrelaxed}(a), one can see a density peaking corresponding to a turbulent particle pinch in the ITG regime. In contrast for the KBM turbulence, the particle flux is outward, see Fig.~\ref{asdex:densrelaxed}(b). Note that there is a weak particle pinch also for the large-aspect ratio tokamak case in the ITG regime. It can be barely seen in Fig.~\ref{dens_relaxed_qa1.9}(a). The particle flux reverses its sign in the KBM regime in the both tokamak geometries considered here. In future work, the question of the particle fluxes driven by the electromagnetic turbulence will be addressed in more detail. 

%
%
\section{Tearing instability and electromagnetic turbulence} \label{Tearing}
%
%
\begin{figure}
	\centering
	\includegraphics[width=0.98\textwidth]{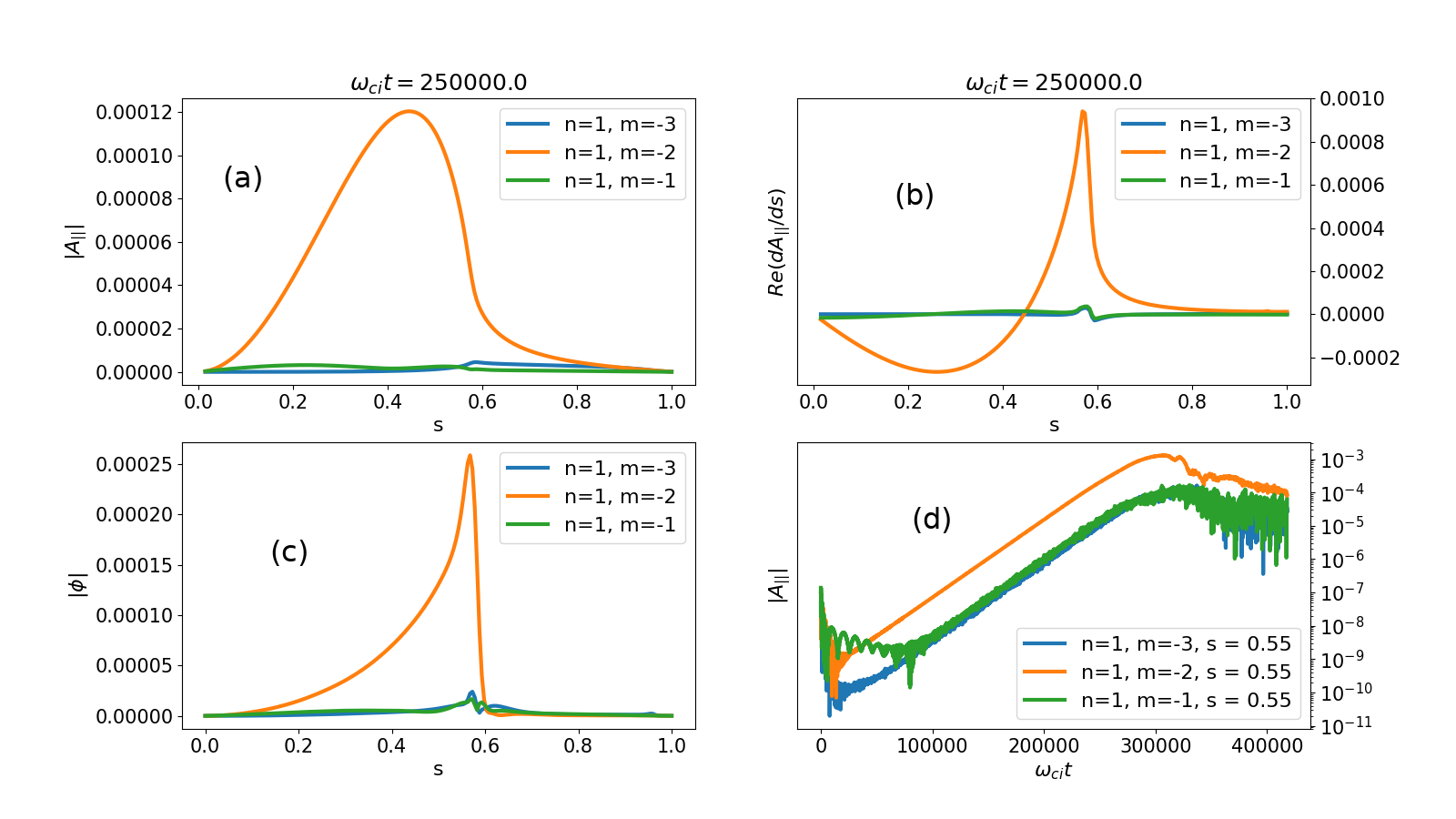}
	\caption{Nonlinear tearing mode. (a) The parallel magnetic potential has a typical structure of the tearing mode instability. (b) The radial derivative of the parallel magnetic potential which is closely related to the tearing instability parameter $\Delta^{\prime}$. It has a distinct spike at the resonant flux surface. (c) The electrostatic potential has a maximum at the resonant flux surface. It's structure is typical for the tearing mode instability. (d) The temporal evolution of the parallel magnetic potential at the $s=0.55$ flux surface. The mode clearly enters the nonlinear phase with the tearing Fourier harmonic $n=1$, $m=-2$ being dominant.}
	\label{tear:phi_mn_nice}
\end{figure}
We consider a tokamak with circular cross-sections, the aspect ratio $A = 10$, the machine size $L_x = 200$, and the safety factor \cite{HornsbyPoP15,Wesson_book}:
\be
q = q_{{\rm a}} \frac{\rho^2/r_{{\rm a}}^2}{1 - (1 - \rho^2/r_{{\rm a}}^2)^{\nu+1}}
\ee
with $\rho$ the radial coordinate, $r_{{\rm a}}$ the minor radius, $q_{{\rm a}} = 3.5$, and $\nu = 1$. This safety factor profile has a $q = 2$ resonance at $s = 0.58$ and is unstable with respect to the tearing mode. In gyrokinetic simulations, the tearing instability drive is included via the shifted Maxwellian distribution function for the electrons with the parallel-velocity shift determined by the ambient parallel current, similar to Ref.~\cite{HornsbyPoP15}. 

In Fig.~\ref{tear:phi_mn_nice}(d), the evolution of the tearing instability is shown for the parallel magnetic potential at the $s=0.55$ flux surface. One sees that the mode enters the nonlinear phase. The simulation is quite expensive computationally because of a relatively low growth rate of the collisionless tearing instability and the physical importance of the collisionless electron skin depth which is resolved in the simulations. The reduced mass ratio $m_{{\rm i}}/m_{{\rm e}} = 200$ and $\beta = 0.4\%$ are used. The mode considered in Fig.~\ref{tear:phi_mn_nice} is driven only by the profile of the ambient parallel current. The plasma temperature and density profiles are flat. The snapshots of the $n=1$ harmonic of the perturbation are shown in Fig.~\ref{tear:phi_mn_nice} for the linear stage at $\omega_{{\rm c}i} t = 0.25 \times 10^6$: (a) the parallel magnetic potential and (c) the electrostatic potential. One sees that a typical tearing mode structures has developed. In Fig.~\ref{tear:phi_mn_nice}(b), the radial derivative of the parallel magnetic potential is plotted. This quantity is closely related to the tearing instability parameter $\Delta^{\prime}$. As expected, it has a sharp maximum at the resonant flux surface.
 
\begin{figure}
	\centering
	\includegraphics[width=0.32\textwidth]{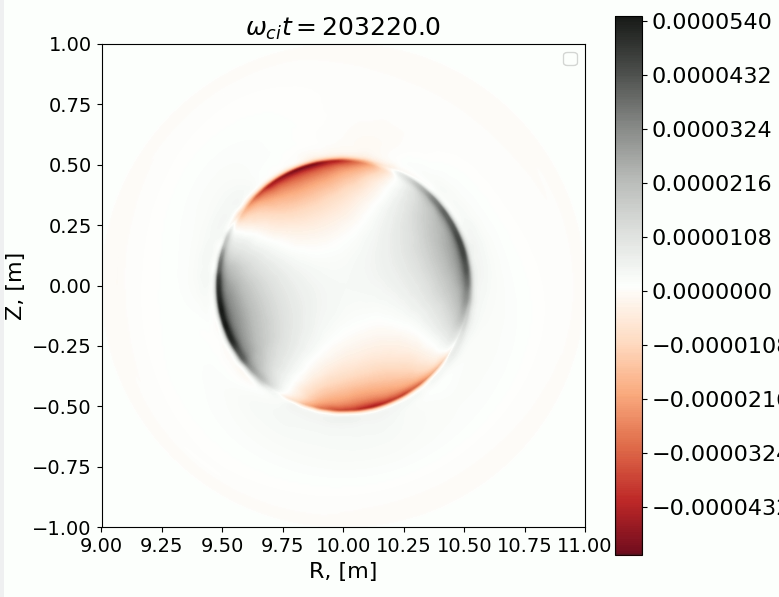}
	\includegraphics[width=0.32\textwidth]{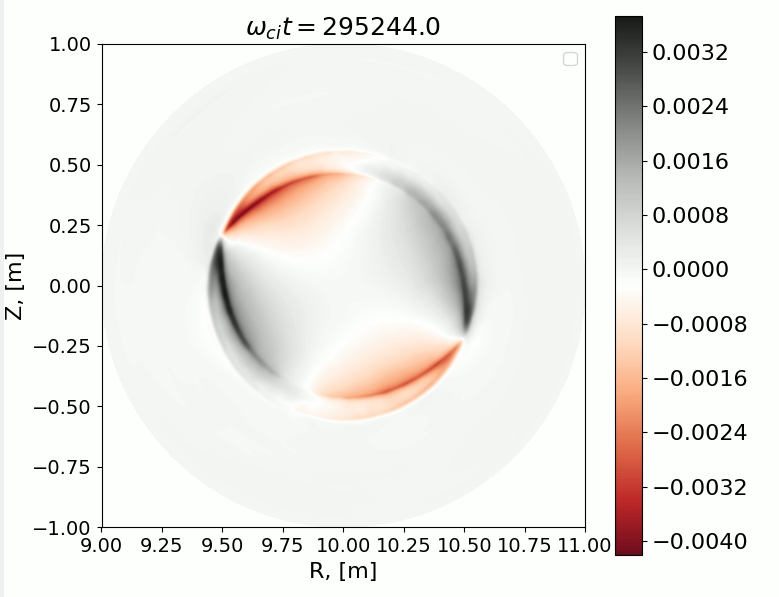}
	\includegraphics[width=0.32\textwidth]{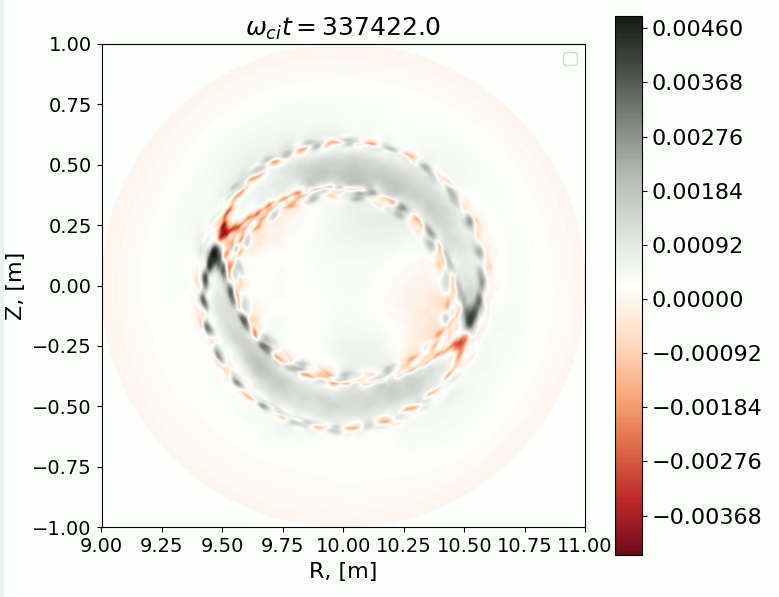}
	\caption{Evolution of the electrostatic potential in tokamak plasmas during the collisionless tearing instability. Flat plasma temperature and density are considered. One sees how an island develops in time with the X-points clearly visible.}
	\label{tear:n0-30_potsc}
\end{figure}
\begin{figure}
	\centering
    \includegraphics[width=0.32\textwidth]{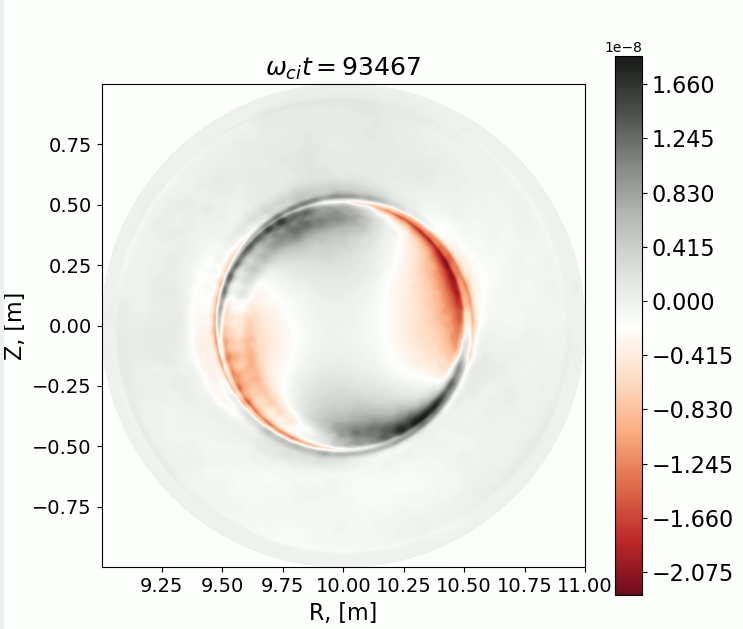}
    \includegraphics[width=0.32\textwidth]{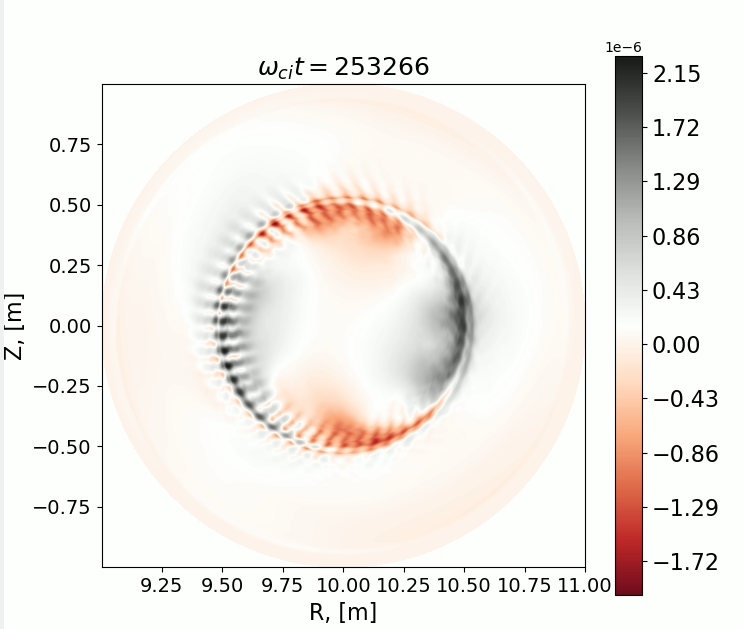}
    \includegraphics[width=0.32\textwidth]{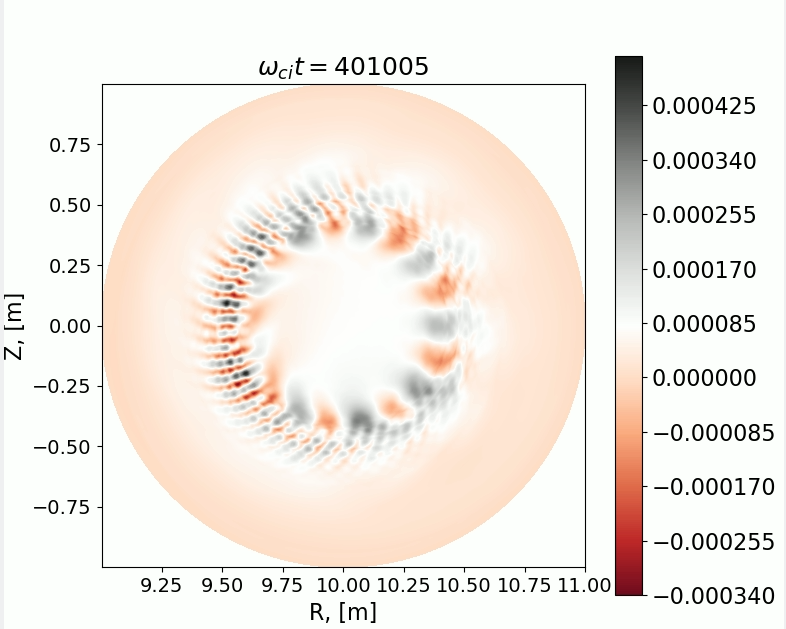}
	\caption{Evolution of the electrostatic potential in tearing-unstable tokamak plasma with finite temperature and density gradients. One sees that the micro-turbulence is excited in addition to the tearing component. Also, a global mode, possibly of an Alfv\'enic nature, is excited at later times.}
	\label{tear:n0-30_kT0.5_potsc}
\end{figure}
In Fig.~\ref{tear:n0-30_potsc}, the evolution of the electrostatic potential is shown in the poloidal cross-section which includes all Fourier harmonics. One can see the tearing mode linearly excited and growing into the nonlinear island structure. The X-points are clearly visible in the nonlinear stage. Toroidal mode numbers $-30 \le n \le 30$ are kept in this simulation and the diagonal poloidal filter is used with the half-width $\Delta m = 5$. 
For comparison, evolution of the electrostatic potential in the case of finite temperature and density gradients is shown in Fig.~\ref{tear:n0-30_kT0.5_potsc}. Here, the plasma profiles are given by Eqs.~(\ref{dens_prof}) and (\ref{temp_prof}) with $s_0 = 0.5$, $\kappa_{{\rm n}} = 0.1$, $\kappa_{{\rm T}} = 0.5$, and $\Delta_{{\rm T}} = \Delta_{{\rm n}} = 0.2$. One sees that the microturbulence develops in addition to the tearing mode excited in the early phase of the simulations via the choice of the initial weights of the markers. At the later stage, a global mode, possibly of an Alfv\'enic nature, appears. We will study this case in detail in our future work. Here, we just note that it could give an example of a self-consistent combination of the MHD tearing instability,  microturbulence, and, eventually, an Alfv\'enic mode and a zonal flow. Such interlinked physical systems \cite{tear_Choi} can be addressed only within a global nonlinear framework and represent the main target of our work.    

\section{Alfv\'en Eigenmodes in the presence of fast ions and electromagnetic turbulence} \label{Fast_ions}
%
%
\begin{figure}
	\centering
	\includegraphics[width=0.32\textwidth]{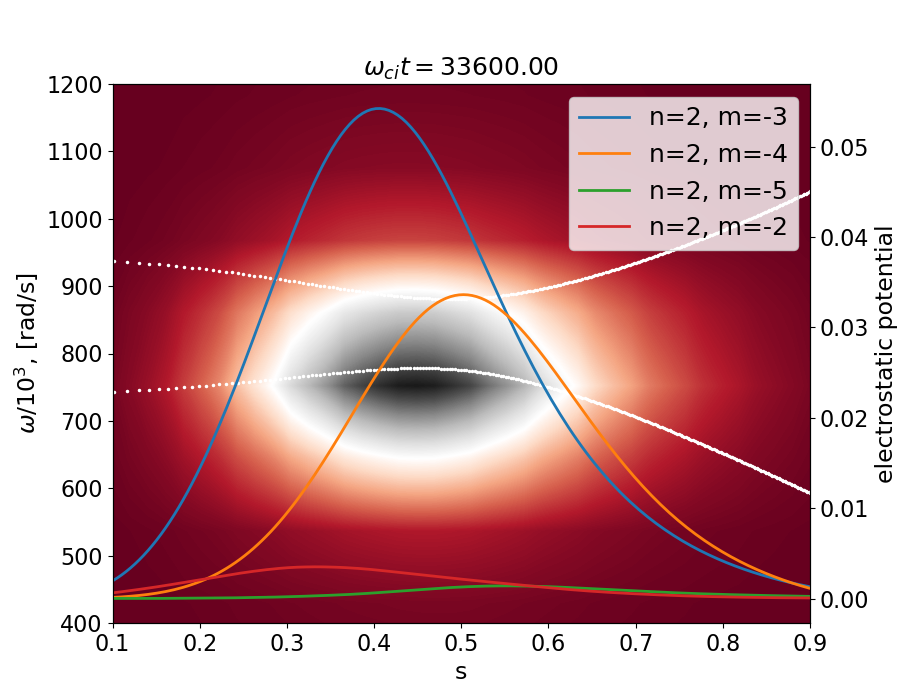}
	\includegraphics[width=0.32\textwidth]{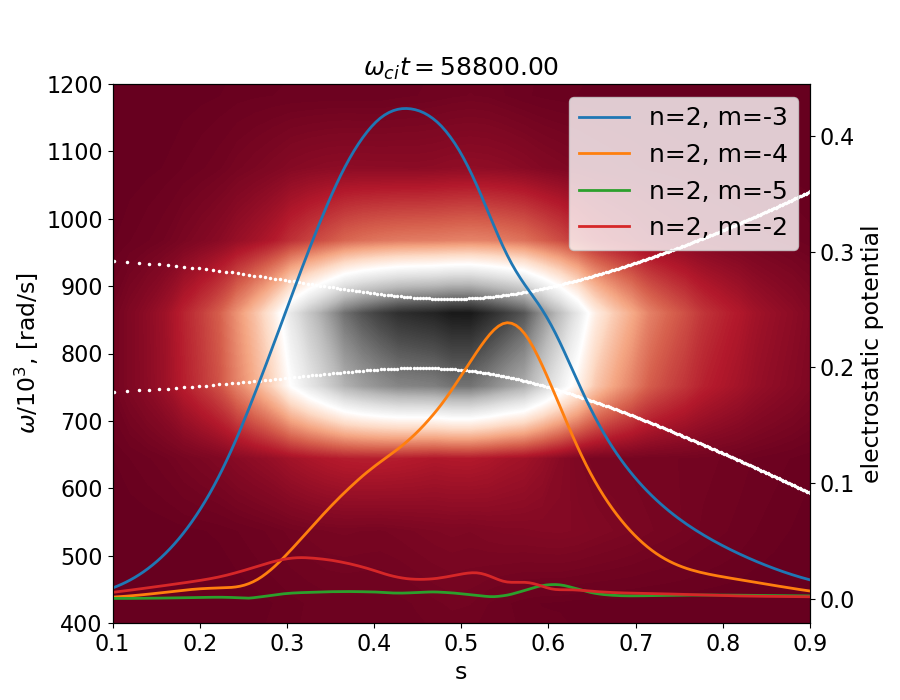}
	\includegraphics[width=0.32\textwidth]{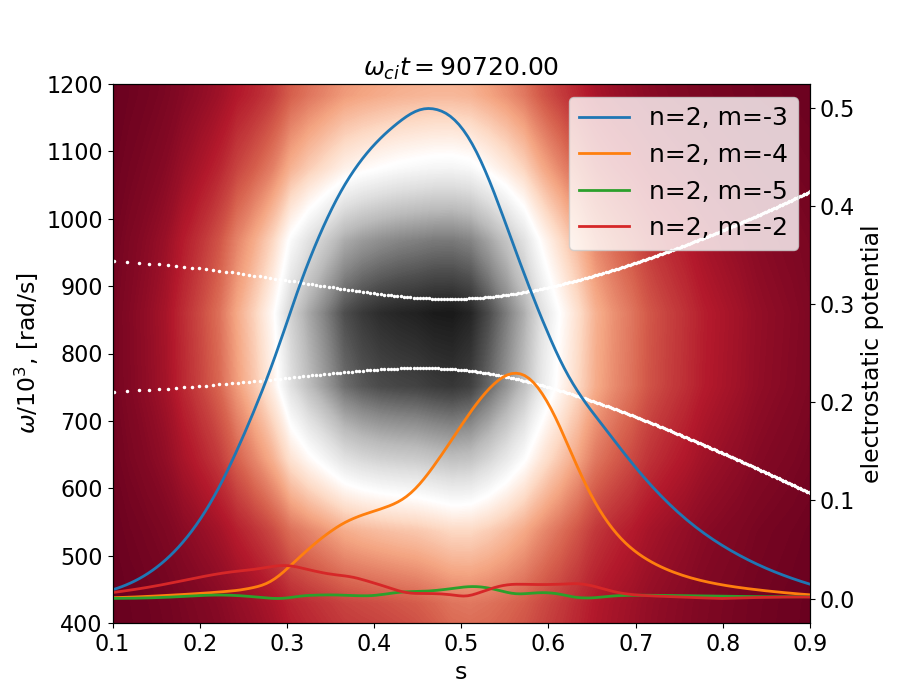}
	\caption{Evolution of the TAE $n=2$ frequency for the case of flat bulk plasma temperature. Here, the windowed Fourier transform of the electrostatic potential is shown for the toroildal mode number $n = 2$ and the poloidal mode numbers $-5 \le m \le -2$. One sees that the frequency of the mode remains at the toroidicity-induced gap. The continuum is computed using the slow-sound approximation.}
	\label{itpa:SAWkT0.0}
\end{figure}

\begin{figure}
	\centering
	\includegraphics[width=0.32\textwidth]{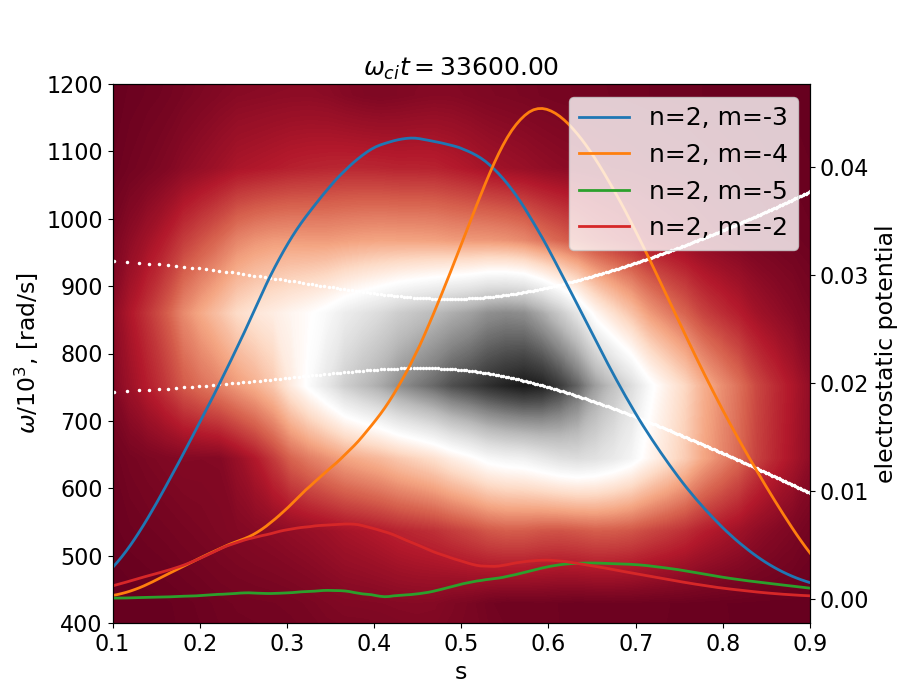}
	\includegraphics[width=0.32\textwidth]{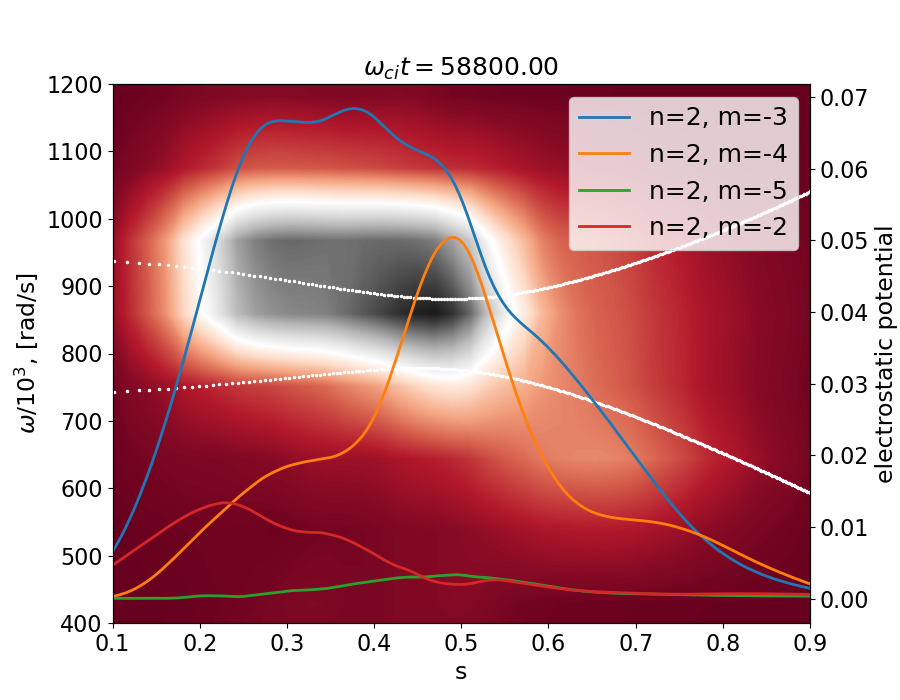}
	\includegraphics[width=0.32\textwidth]{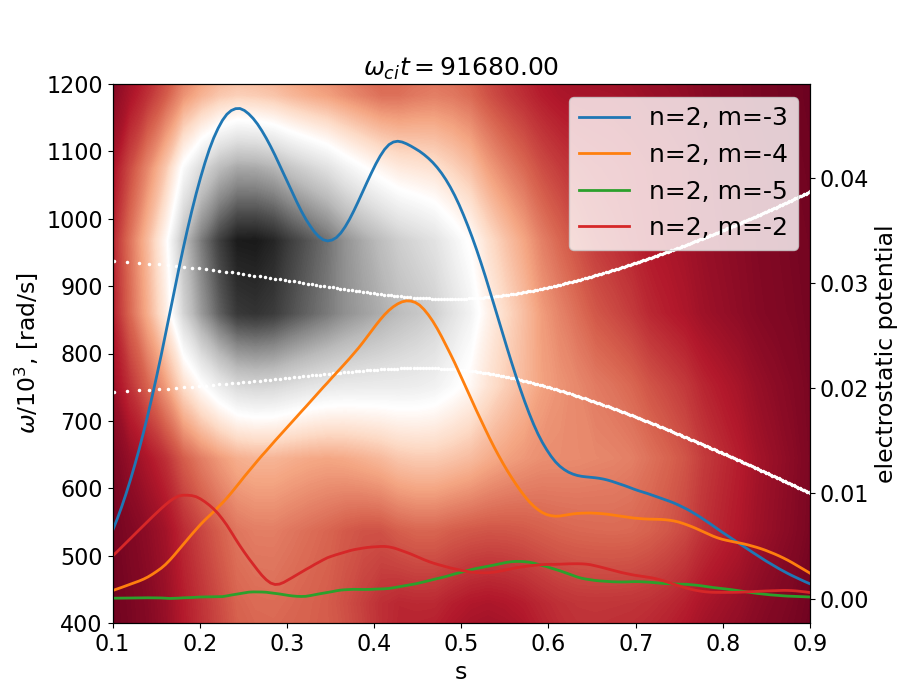}
	\caption{Simulations using the same parameters as in Fig.~\ref{itpa:SAWkT0.0} except the bulk-plasma temperature which has now a weak gradient $\kappa_{{\rm T}} = 0.4$, see Eq.~(\ref{temp_prof4}). One sees that the frequency evolves along the continuum branch. The same quantity is shown as in Fig.~\ref{itpa:SAWkT0.0} (the Fourier transform of the electrostatic potential).}
	\label{itpa:SAWkT0.4}
\end{figure}

\begin{figure}
	\centering
	\includegraphics[width=0.47\textwidth]{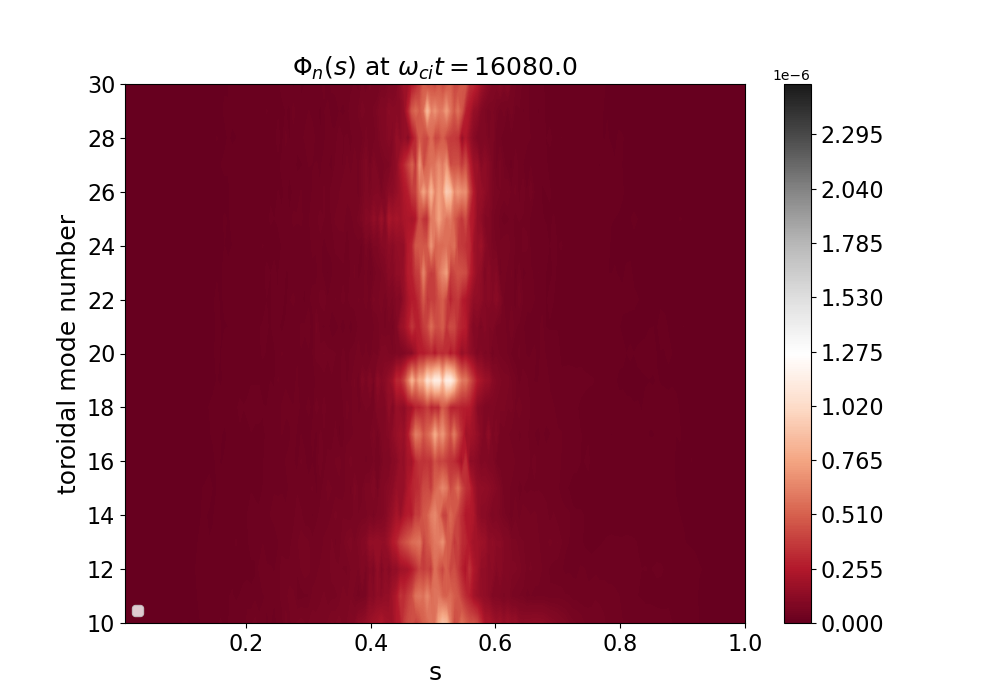}
	\includegraphics[width=0.47\textwidth]{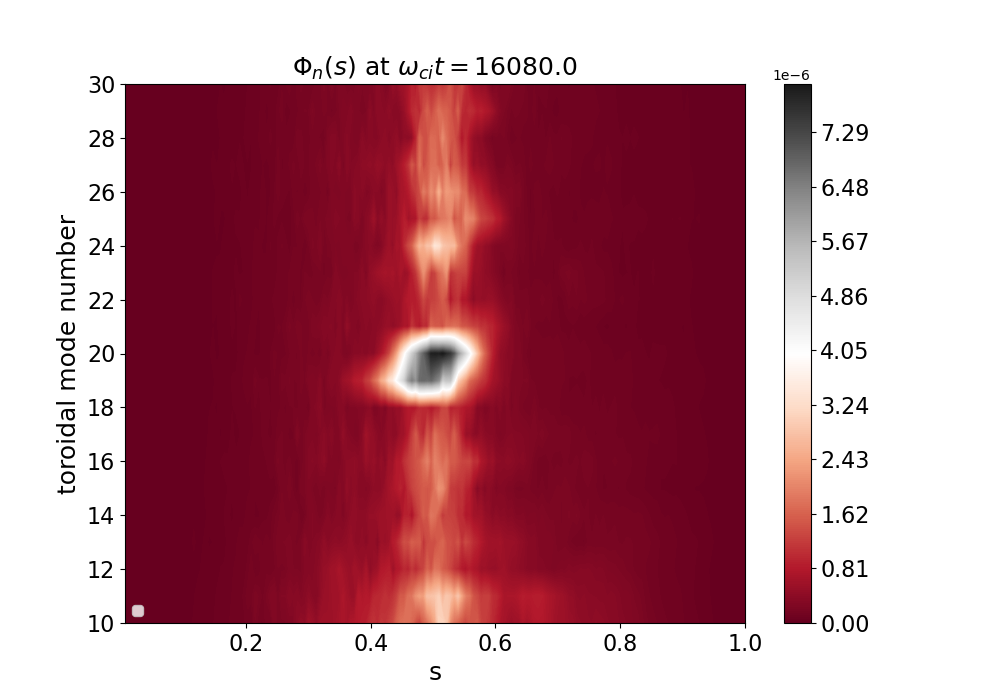}
	\caption{Electrostatic potential plotted as a function of the flux surface label $s$ and the toroidal mode number $n$. Only the relevant part of the spectrum is shown. One sees that there is a mode activity in the case with the finite temperature gradient (b). In contrast, the case of the flat temperature gradient (a) does not have the mode activity. A consequence of the mode activity in (b) is a higher amplitude of the perturbation in the nonlinear phase which can be related to the characteristic frequency evolution observed in Fig.~\ref{itpa:SAWkT0.4}.}
	\label{itpa:potsn}
\end{figure}

In this Section, we address the issue of coupling between the turbulence and Alfv\'en Eigenmodes destabilized by the fast ions. We consider a circular cross-section tokamak with the aspect ratio $A = 10$ and the machine size $L_x = 350$. The magnetic geometry is that of the ITPA benchmark \cite{Koenies_itpa_NF,Mishchenko_fast}. The simulation includes the toroidal Fourier modes $-40 \le n \le 40$ and all the poloidal modes inside a diagonal filter with the half-width $\Delta m = 5$. All particle species (bulk ions, electrons, and fast ions) are kinetic and nonlinear. The reduced mass ratio $m_{{\rm i}}/m_{{\rm e}} = 200$ and $\beta = 1.72\%$ are used. For the fast particles, Deuterium ions are selected with the temperature $T_{{\rm f}}/T_{{\rm i}} = 40$ and the density $n_{{\rm f}}/n_{{\rm i}} = 0.005$. Here, $T_{{\rm i}}$ is the bulk-ion temperature and $n_{{\rm i}}$ is the bulk-ion density. The Maxwell distribution function is chosen for the fast ions. The fast-ion temperature profile is flat, the fast-ion density profile is given by Eq.~(\ref{dens_prof}) with $\kappa_{{\rm n}} = 3.33$, $s_0 = 0.5$, and $\Delta_{{\rm n}} = 0.2$.

In Fig.~\ref{itpa:SAWkT0.0}, the frequency evolution is shown for the case of a flat bulk-ion profile for the toroidal Fourier harmonic $n=2$. It corresponds to the dominant TAE instability for the parameters considered. There are other TAEs, for example at the toroidal mode number $n = 6$, which are however less unstable. One can see that the frequency remains at the toroidicity-induced gap for all times of the nonlinear evolution. The shear-Alfv\'en continuum is computed using the CONTI code \cite{conti_2010,conti_2020} within the slow-sound approximation. 

In contrast, the frequency of the same TAE evolves along the continuum branch when the bulk-plasma species has a temperature gradient, see Fig.~\ref{itpa:SAWkT0.4}. Here, we apply a rather shallow bulk-plasma temperature profile given by the expression:
\be
\label{temp_prof4}
T_{0s}(\rho)/T_{0s}(\rho_0) = \exp\left[-\frac{\kappa_{{\rm T}} \Delta_{{\rm T}}}{2} {\rm ln}\left[ \frac{{\rm cosh}\left(\frac{\rho - \rho_0 + \Delta\rho}{\Delta_{{\rm T}}}\right)}{{\rm cosh}\left(\frac{\rho - \rho_0 - \Delta\rho}{\Delta_{{\rm T}}}\right)} \right]\right]
\ee
with $\rho$ the minor radius, $\kappa_{{\rm T}} = 0.4$, $\Delta_{{\rm T}} = 0.04$, $\rho_0 = 0.5$, and $\Delta\rho = 0.4$. 
All other physical and computational parameters remain unchanged.
A possible explanation for this frequency evolution is a higher amplitude of the Fourier harmonics corresponding to the drift instabilities. An indication for such instabilities excited by the temperature profile considered can be seen in Fig.~\ref{itpa:potsn}. Here, the elecrostatic potential is plotted as a function of the flux surface label and the toroidal mode number. One can see a coherent instability developing at $n \approx 20$ when the bulk-plasma temperature has a gradient. 
%
Clearly, more work needs to be done in future in order to fully understand the physical mechanism behind these observations. 

%
%
%
\section{Stellarator turbulence} \label{Stellarators}
\begin{figure}
	\centering
	\includegraphics[width=0.59\textwidth]{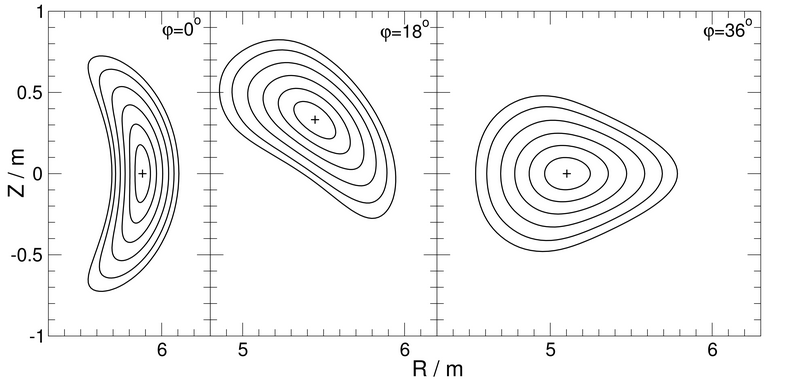}
	\includegraphics[width=0.4\textwidth]{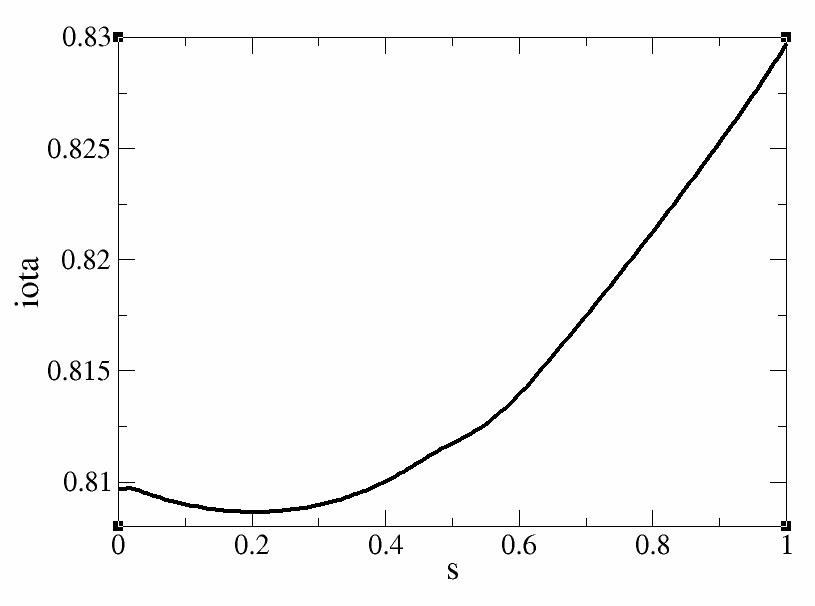}
	\caption{The characteristic cross-sections at several toroidal angles and the rotational transform (iota) of the W7-X configuration considered. The normalized toroidal flux $s$ is used as the radial coordinate for the rotational transform.}
	\label{w7x:fluxsurfs}
\end{figure}
\begin{figure}
	\centering
	\includegraphics[width=0.47\textwidth]{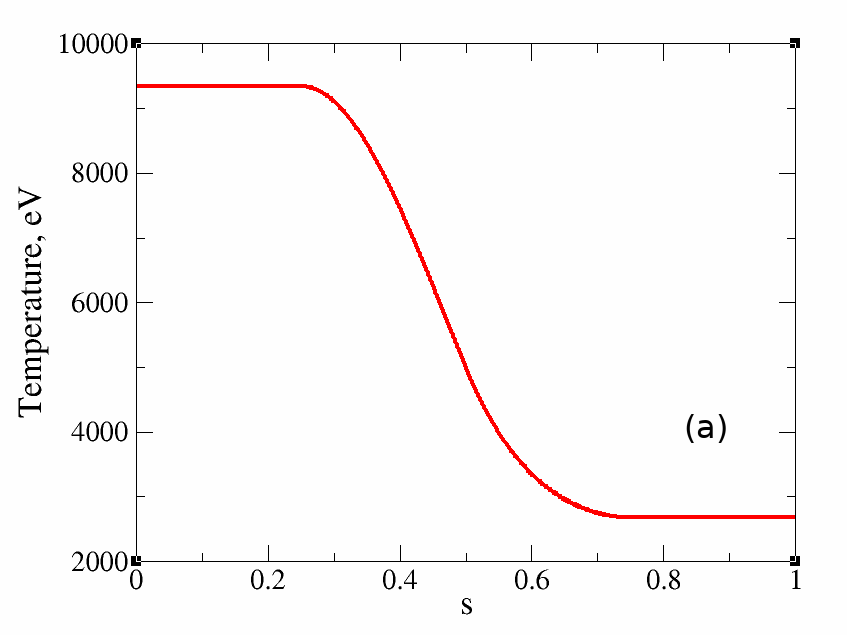}
	\includegraphics[width=0.47\textwidth]{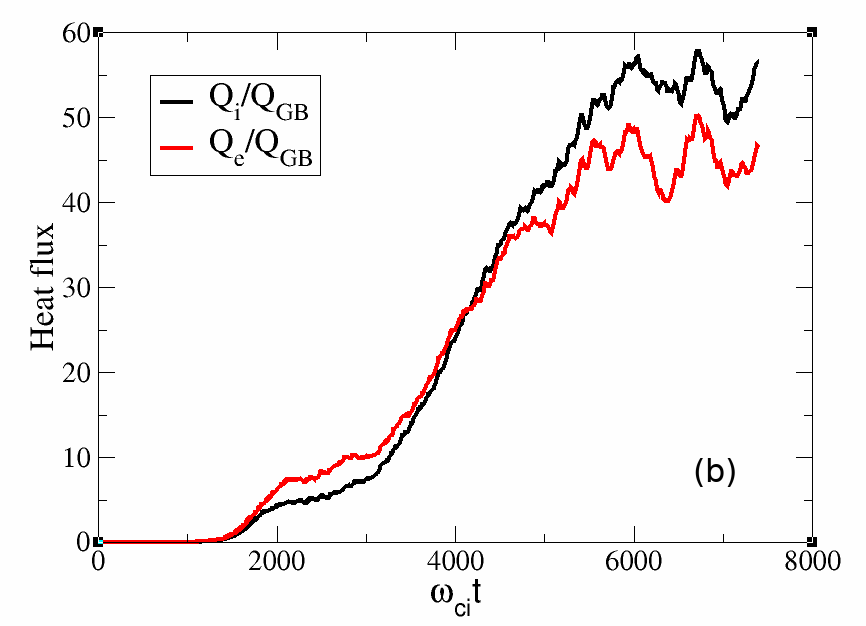}
	\caption{The plasma temperature (a) and the resulting turbulent heat flux (b) in W7-X. The normalized toroidal flux $s$ is used as the radial coordinate.}
	\label{w7x:fluxes}
\end{figure}
\begin{figure}
	\centering
	\includegraphics[width=0.32\textwidth]{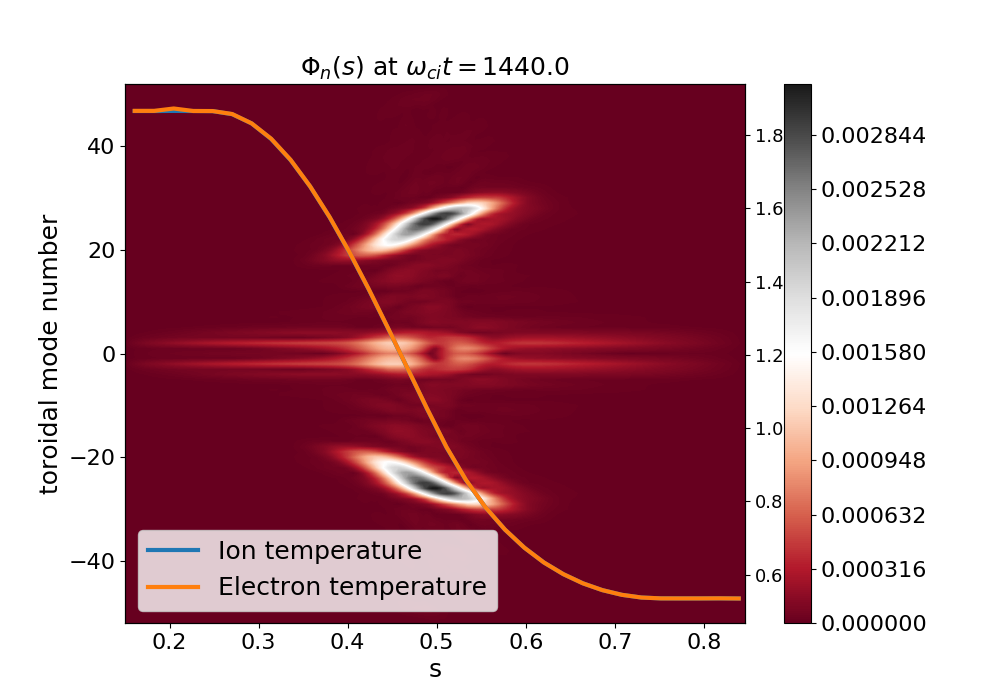}
	\includegraphics[width=0.32\textwidth]{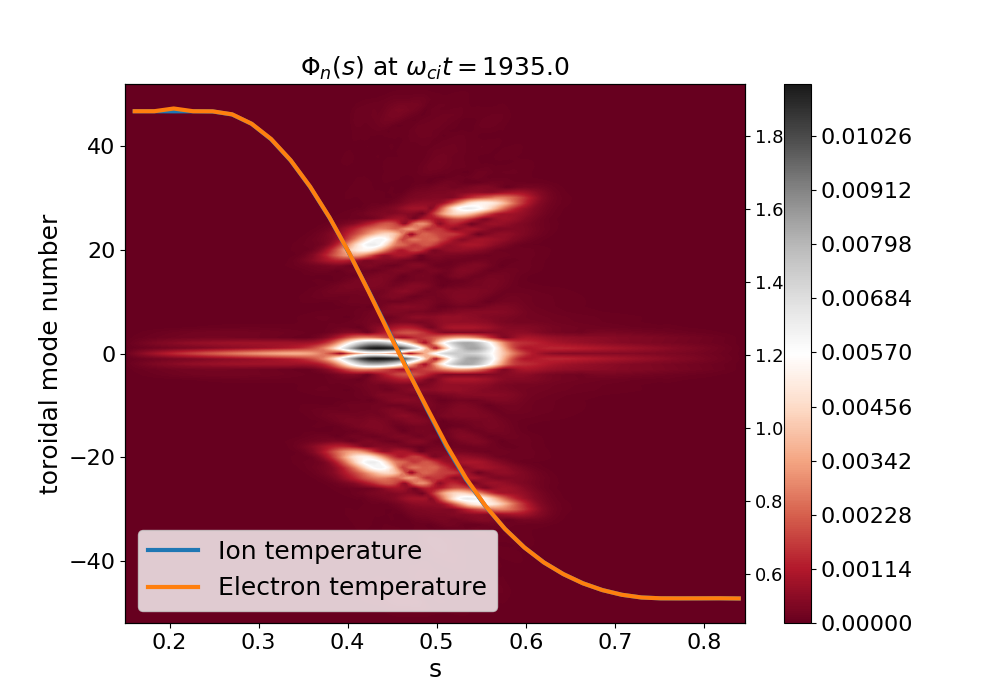}
	\includegraphics[width=0.32\textwidth]{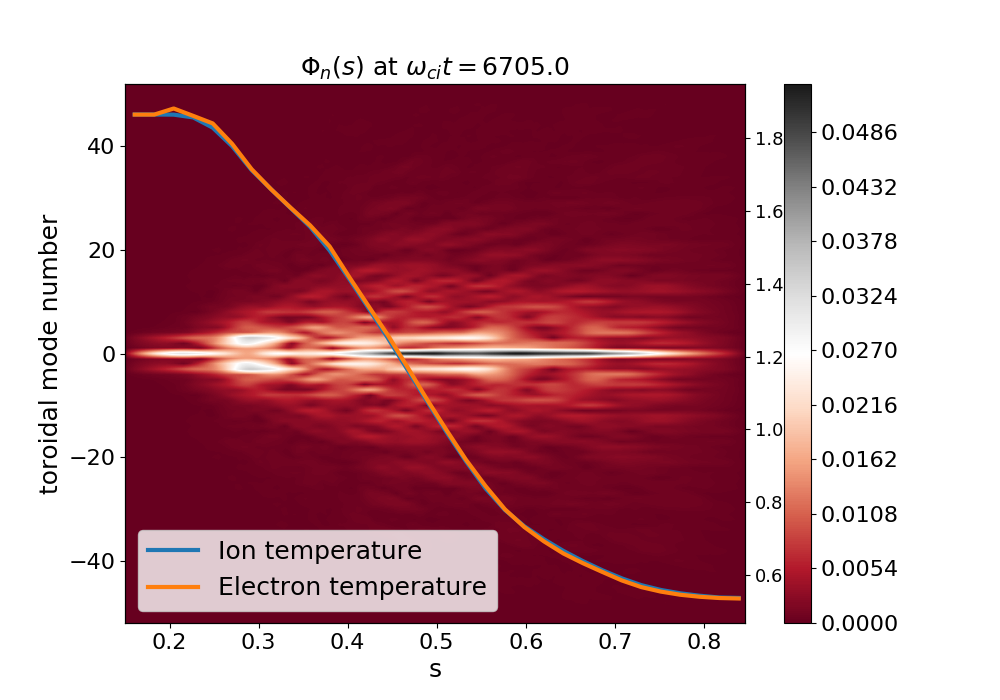}
	\caption{Evolution of the electrostatic potential in the W7-X as a function of the flux label and toroidal mode number. One can see how the linear instability evolves generating the low-mode-number components (including the zonal flows) and modifying the ambient temperature profile in the nonlinear phase.}
	\label{w7x:potsn}
\end{figure}
\begin{figure}
	\centering
	\includegraphics[width=0.32\textwidth]{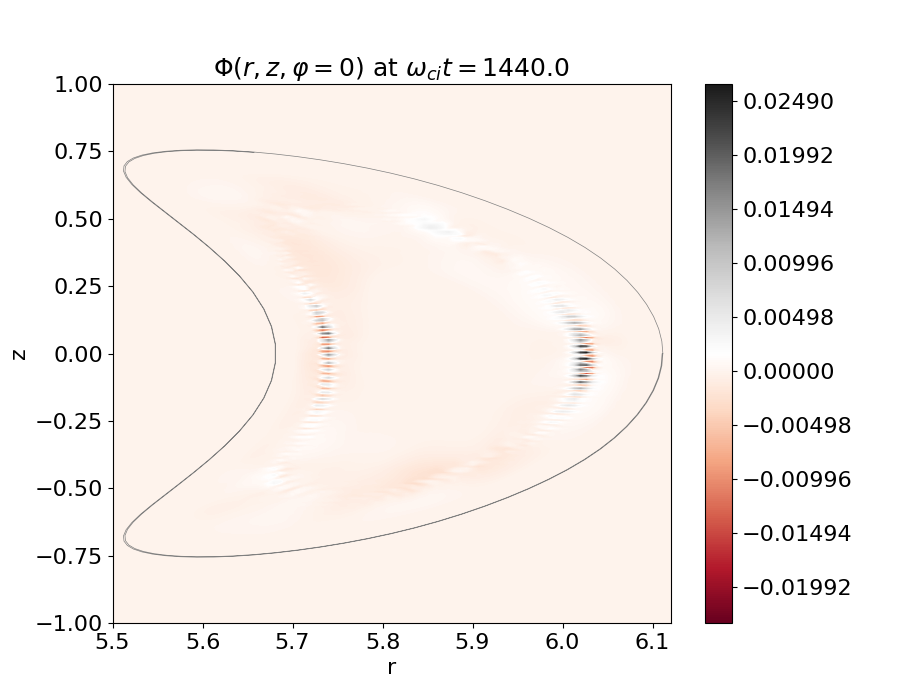}
	\includegraphics[width=0.32\textwidth]{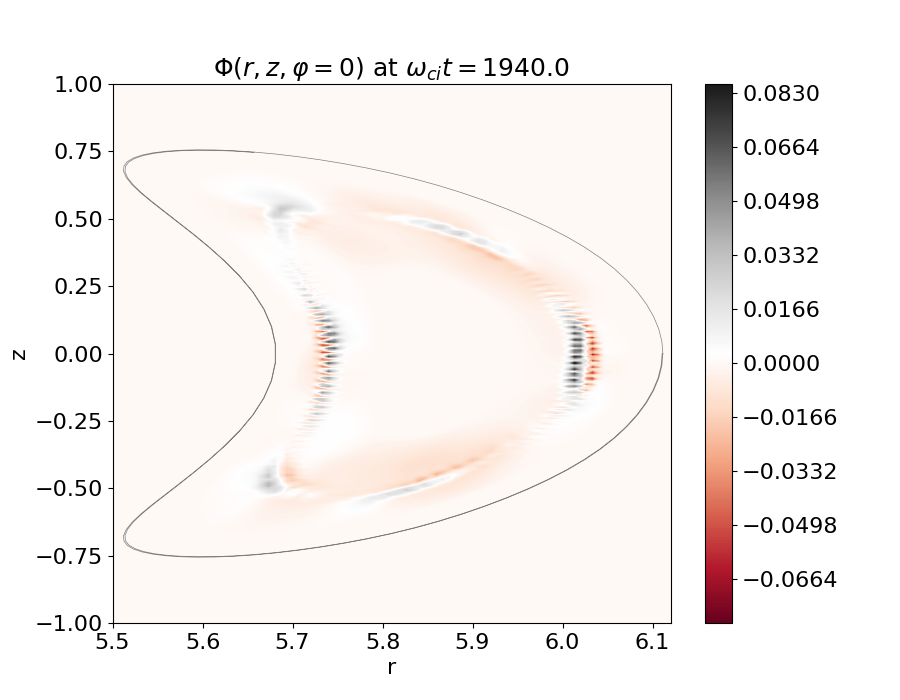}
	\includegraphics[width=0.32\textwidth]{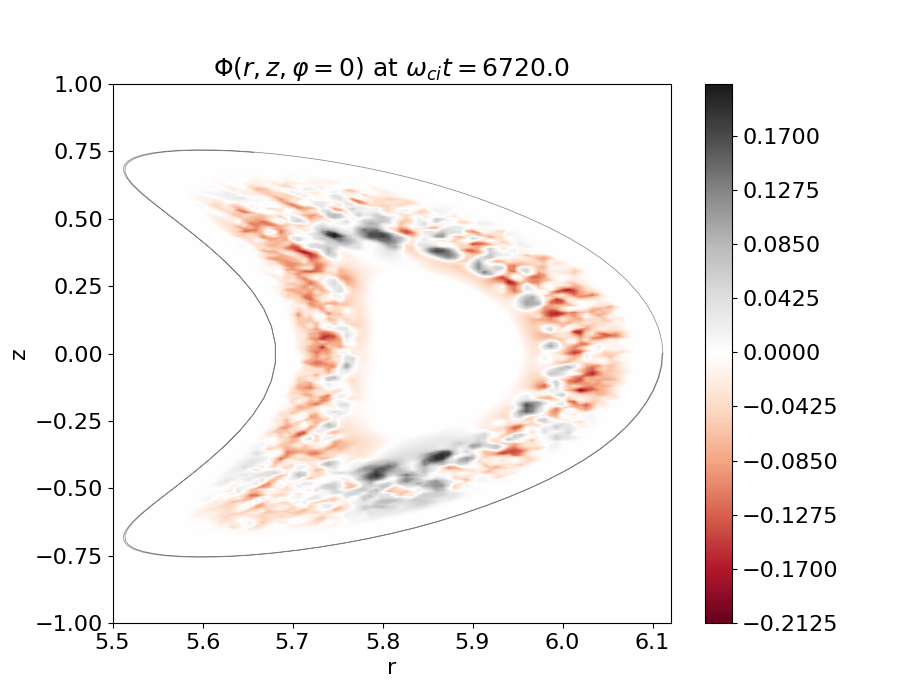}
	\caption{Evolution of electrostatic potential in the W7-X shown in the poloidal cross-section corresponding to the toroidal angle $\varphi=0$.}
	\label{w7x:potrz}
\end{figure}
In this Section, we apply the EUTERPE code to study the electromagnetic turbulence in a stellarator plasma.
We consider a Wendelstein 7-X \cite{wolf:pop:2019}\ 
configuration falling into the class of variants which exploit
the rotational transform $\iota=5/6$ and the corresponding edge islands for divertor operation.
Furthermore, the specific case chosen here is characterized by a very high 
magnetic mirror field,
on the magnetic axis $(B_{\sf max} - B_{\sf min})/(2 <B>) \approx 0.25$,
which has to be compared to the magnetic mirror of $0.1$ in the Wendelstein 7-X standard
case.
On the one hand, this magnetic mirror leads to a small bootstrap current, 
on the other hand, it prevents the formation of a vacuum-field magnetic well. 
In fact, this configuration even possesses a vacuum-field magnetic hill, 
which means that, when going from the magnetic axis to the plasma boundary,
the enclosed volume grows more strongly than the enclosed magnetic flux. 
Ideal MHD theory shows that such magnetic configurations are plagued 
by the lack of ideal MHD stability. In Ref.~\cite{cas:nf:2016}, a similar W7-X variant was 
found to be unstable against low-mode-number ideal MHD modes at even low plasma pressure.

For the configuration studied here a free-boundary plasma equilibrium was computed using
the ideal MHD equilibrium code VMEC \cite{hirshman:1986}. For a plasma volume of 
$V_{\sf plasma}=25~{\sf m}^3$,
a uniform number density of $n_0= 1.4 \times 10^{19}~{\sf m}^{-3}$ 
and the temperature profile shown in Fig.~\ref{w7x:fluxes}(a) correspond to $\beta=1.52\%$
at the center of the simulation volume, i.~e.~at the normalized toroidal flux $s=0.5$. The characteristic cross-sections illustrating the 
non-axisymmetric geometry of the stellarator and the rotational transform are shown in Fig.~\ref{w7x:fluxsurfs}.
%
The reduced mass ratio $m_{{\rm i}}/m_{{\rm e}} = 200$ is used. 
%
The resulting turbulent evolution of the heat flux is shown in Fig.~\ref{w7x:fluxes}(b). One can see that the simulation clearly enters the nonlinear phase. 

The electrostatic potential at different times is shown in Fig.~\ref{w7x:potsn} as a a function of the flux surface label and the toroidal mode number. Here, one can see that a linear instability develops at early times. The low-mode-number part of the spectrum (which includes the zonal flow) is excited when the turbulence evolves. At later times, the temperature profile is modified via the turbulent relaxation.
In Fig.~\ref{w7x:potrz}, the electrostatic potential is shown in the poloidal cross-section corresponding to the toroidal angle $\varphi = 0$. One sees that the perturbation covers only a small fraction of the poloidal angles in the linear regime. It spreads over the whole poloidal domain in the nonlinear phase.
In future, we will study global electromagnetic turbulence in stellarator plasma in detail. 
%
%
\section{Conclusions}  \label{Conclusions}
In this paper, we have considered the electromagnetic turbulence in tokamak plasmas. The ITG-KBM transition has been identified and the relaxation of the profiles in the case of the KBM turbulence has been observed. For the large-aspect-ratio tokamak, the mode saturation in the KBM regime appears to be due to the flattening of the temperature profile caused by an outward transport of the turbulent finger-like structures. We have considered the electromagnetic turbulence in a down-scaled ASDEX-Upgrade plasma where a similar ITG-KBM transition has been observed as well. The profile relaxation is weaker in the ASDEX-Upgrade for the parameters considered. The multiscale physics has been addressed coupling the electromagnetic turbulence to the collisionless tearing instability and Alfv\'enic modes destabilized by the fast ions. Electromagnetic turbulence has also been addressed in the non-axisymmetric stellarator geometry. 

A typical computational effort for the cases considered here requires large jobs on Marconi (128 nodes for many days of the simulation duration) or Marconi100 (64 nodes). In future, exascale computing systems will be needed for the reactor-size turbulence simulations and for massive parameter scans using the global codes. The spatial resolution can also become an issue when the collisionless reconnection is of importance requiring the collisioneless electron skin depth to be resolved. Extending the simulations to the realistic mass ratio does not normally pose a problem. Of course, such simulations are computationally more expensive because of a smaller time step needed. This is the main reason to use a reduced mass ratio throughout this paper. A number of simulations at the realistic mass ratio has also been performed for various scenarios and will be reported elsewhere. Since the collisionless electron skin depth is affected when the electron mass is decreased, the radial resolution requirements can become more challenging for the realistic mass ratio when the tearing mode is present.

The main goal of this paper has been to present a set of the characteristic physics applications which can be used as a starting point for future more detailed numerical experiments with ORB5 and EUTERPE. Here, we demonstrate that the simulations of such kind are possible using existing global gyrokinetic particle-in-cell codes on the available HPC systems. 
%
%
\section*{Acknowledgments}
This
work has been carried out within the framework of the EUROfusion Consortium, funded by the European Union via the Euratom Research and
Training Programme (Grant Agreement No 101052200 – EUROfusion). Views and opinions expressed are however those of the authors only and do not necessarily reflect those of the European Union or the European Commission. Neither the European Union nor the European Commission can be held responsible for them. 
This work was supported in part by the Swiss National Science Foundation. 
Simulations presented in this work were performed on the MARCONI FUSION HPC system at CINECA. 
We acknowledge PRACE for awarding us access to Marconi100 at CINECA, Italy.
We acknowledge PRACE for awarding us access to Joliot-Curie at GENCI@CEA, France.

\section*{References}
\bibliographystyle{unsrt} 
\bibliography{iter_turb.bib}


\end{document}